\documentclass[a4paper,12pt]{article}
\usepackage{amsmath}
\usepackage{amssymb}

 \newread\testifexists
\def\GetIfExists #1 {\immediate\openin\testifexists=#1
    \ifeof\testifexists\immediate\closein\testifexists\else
    \immediate\closein\testifexists\input #1\fi}

\immediate\openin\testifexists=e
    \ifeof\testifexists\immediate\closein\testifexists\else
    \immediate\closein\testifexists\input e\fipsf

\begin{document}

\title{ Polarization, CP Asymmetry and Branching ratios in $B \to K^{\ast} K^{\ast}$ with Perturbative QCD approach}
\author{Jin Zhu$^{a,b,c}$\footnote {zhujin@mail.ihep.ac.cn} \, Yue-Long  Shen$^{a,b,c}$\footnote {shenyl@mail.ihep.ac.cn}
\, Cai-Dian L\"u$^{a,b}$ \,      \\
 {\it \small $a$ CCAST (World Laboratory), P.O. Box 8730,
   Beijing 100080, P.R. China}\\
{\it \small $b$  Institute of High Energy Physics, CAS, P.O.Box
918(4) } {\it \small 100049, P.R. China}\footnote {Mailing
address}\\
 {\it \small $c$  Graduate School of Chinese Academy of
Science, P.R. China }} \maketitle

\begin{abstract}
We study the charmless rare decays $B \to K^{\ast} K^{\ast}$
within the Perturbative QCD picture. We calculate not only
factorizable and non-factorizable diagrams, but also annihilation
ones. Our predictions are the following: The longitudinal
polarization fraction vary from $75\%$ to $99\%$ depending on
channels, the branching ratios are of order $10^{-7}$ for
$B^0(\bar{B}^0) \to K^{\ast0}\bar{K}^{\ast0}$ and $B^{\pm} \to
K^{\ast\pm} \bar{K}^{\ast0}(K^{\ast0})$, much bigger than that for
$B^0(\bar{B}^0) \to K^{\ast+}K^{\ast-}$($10^{-8}$). The direct CP
asymmetry in $B^{\pm} \to K^{\ast\pm} \bar{K}^{\ast0}(K^{\ast0})$
and $B^0(\bar{B}^0) \to K^{\ast+}K^{\ast-}$ is about $-15\%$ and
$-65\%$ if we choose $\alpha (\varphi_2)$ as $95^{\circ}$. There's
no direct CPV in $B^0(\bar{B}^0) \to K^{\ast0}\bar{K}^{\ast0}$
decays because of the pure $b \to d$ penguin topology. Our
predictions will be tested in the future B experiments.
\end{abstract}

\section{Introduction}
\hspace*{\parindent} Exclusive B meson decays, especially $B \to
VV$ modes, have aroused more and more interest for both theorists
and experimenters. Since it offers an attractive opportunity to
get a deep insight into the flavor structure of the Standard Model
(SM) and the CP violation parameters. But things are not so easy
due to the Non-perturbative QCD dynamics. Several approaches,
which include factorization approach (FA) \cite{bsw,ali}, QCD
improved factorizations (QCDF ) \cite{bbns,beneke}, Soft-collinear
effective theory (SCET) \cite{scet} and Perturbative QCD (PQCD)
\cite{hnli,keum,pipi} approach, have been developed to solve this
problem. PQCD is based on $k_T$ factorization theorem
\cite{kt1,kt2,kt3} while others are most based on collinear
factorization \cite{colliearfa}. Besides, Sudakov factor and
threshold resummation \cite{kt1,threshold} have been induced in
PQCD to regulate the End-point singularities, so the arbitrary
cutoffs \cite{beneke} are no longer necessary.

   In the PQCD framework the hard amplitudes for various topologies of
diagrams, including factorizable, nonfactorizable and
annihilation, are all six-quark amplitudes, while in FA and QCDF
the leading factorizable diagrams involve four-quark amplitudes.
This difference leads to a different characteristic scale, the
former $\sqrt{\wedge m_B}\ \ (\sim 1.5GeV)$ \cite{keum,pipi} and
the latter $m_B$. Therefore, we get a larger Wilson coefficients
($C_{3-6}$) associated with the QCD penguin in PQCD due to the
evaluation of the renormalization group, this means penguin
diagrams have been enhanced dynamically. Current penguin dominated
modes data, such as $B \to K \pi$ \cite{particle} and $B \to \phi
K$ \cite{exphik}, seem to fit well with the PQCD predictions
\cite{keum,pipi,phik}.

The recent $B \to \phi K^*$ data
\cite{bellephikstar,babarphikstar} reveal a large transverse
polarization fraction, which differs from most theoretical
predictions and is considered as a puzzle. This indicates $B\to
VV$ modes must be more complicated than we think and needs to be
investigated more thoroughly. Motivated by this, we study another
$B\to VV$ mode within the Standard Model (SM). $B\to K^*K^*$
decays, which governed by $B \to K^*$ form factors too, may help
us to know more about the polarization puzzle, as well as the CKM
phase angle $\alpha$ \cite{mearalpha} and new physics. In the
following sections, we will perform $B \to K^{\ast} K^{\ast}$
decays, which have the same topologies with $B \to K K$ \cite{kk},
within the PQCD framework. Our goal is to find out the branching
ratios, CP asymmetries, as well as the polarization fractions.

\section{Framework and power counting}

\hspace*{\parindent} In the $B \to K^{\ast} K^{\ast}$ modes, the B
meson is heavy and sitting at rest. It decays to two light vector
mesons with large momenta. Therefore the $K^{\ast}$ mesons are
moving very fast in the rest frame of B meson. In this case, the
short distance hard process dominates the decay amplitude and
Final State Interaction (FSI) may not be important in most of the
cases, this makes the perturbative QCD applicable. For
$B^0(\bar{B}^0) \to K^{\ast+}K^{\ast-}$ decay, because of its
small branching ratio, FSI may occur through intermediate states
$\rho\rho$ or $K^{*0}\bar K^{*0}$, etc. If future experiments
deviate theoretical prediction largely, it might be an indication
of strong FSI effects. Here we give only the perturbative picture
for experiments to test.

 In PQCD approach, the decay amplitude is factorized into
  the convolution of the mesons'
light-cone wave functions, the hard scattering kernel and the
Wilson coefficients, which stands for the soft, hard and harder
dynamics respectively. The transverse momentum was introduced so
that the endpoint singularity which will break the collinear
factorization is regulated and the large double logarithm term
appears after the integration on the transverse momentum, which is
then resummed into the Sudukov form factor. The formalism can be
written as:
\begin{eqnarray}
{\cal M}\sim\nonumber &&\int
dx_1dx_2dx_3b_1db_1b_2db_2b_3db_3Tr[C(t)\Phi_B(x_1,b_1)\Phi_{K^*}(x_2,b_2)
\Phi_{K^*}(x_3,b_3)\\ &&H(x_i,b_i,t)S_t(x_i)e^{-S(t)}],
\end{eqnarray}
where the $b_i$ is the conjugate space coordinate of the
transverse momentum, which represents the transverse interval of
the meson. $t$ is the largest energy scale in hard function $H$,
while the jet function $S_t(x_i)$ comes from the summation of the
double logarithms $\ln^2x_i$, called threshold resummation
\cite{kt1,threshold}, which becomes large near the endpoint.

    We use the effective Hamiltonian for the
process $B \to K^* K^*$ given by \cite{Buchalla}
\begin{eqnarray}
{\cal
H}_{eff}=\frac{G_F}{\sqrt{2}}\left\{V_u\left[C_1(\mu)O_1(\mu)+C_2(\mu)O_2(\mu)\right]
-V_t\sum_{i=3}^{10}C_i(\mu)O_i^{(q)}(\mu)\right\},
\end{eqnarray}
where the CKM matrix elements $V_u=V_{ud}^*V_{ub}$,
$V_t=V_{td}^*V_{tb}$, $C_i(\mu)$ being the Wilson coefficients,
and the operators
\begin{eqnarray}
\nonumber &&O_1=(\bar{d}_iu_j)_{V-A}(\bar{u}_jb_i)_{V-A},\,
O_2=(\bar{d}_iu_i)_{V-A}(\bar{u}_jb_j)_{V-A}, \\
\nonumber
&&O_3=(\bar{d}_ib_i)_{V-A}\sum\limits_q(\bar{q}_jq_j)_{V-A},\,
O_4=(\bar{d}_ib_j)_{V-A}\sum\limits_q(\bar{q}_jq_i)_{V-A}, \\
\nonumber &&
O_5=(\bar{d}_ib_i)_{V-A}\sum\limits_q(\bar{q}_jq_j)_{V+A},\,
 O_6=(\bar{d}_ib_j)_{V-A}\sum\limits_q(\bar{q}_jq_i)_{V-A},\\ \nonumber
 &&O_7=\frac{3}{2}(\bar{d}_ib_i)_{V-A}\sum\limits_qe_q(\bar{q}_jq_j)_{V+A},\,
 O_8=\frac{3}{2}(\bar{d}_ib_j)_{V-A}\sum\limits_qe_q(\bar{q}_jq_i)_{V+A},
 \\ &&
 O_9=\frac{3}{2}(\bar{d}_ib_i)_{V-A}\sum\limits_qe_q(\bar{q}_jq_j)_{V-A},\,
 O_{10}=\frac{3}{2}(\bar{d}_ib_j)_{V-A}\sum\limits_qe_q(\bar{q}_jq_i)_{V-A}.
 \end{eqnarray}
 i and j stand for $SU(3)$ color indices.

 Now let's analyze these decay channels
topologically. First, it is categorized emission and annihilation
diagrams. Second, each category can be extracted to 4 diagrams,
two factorizable and two nonfactorizable, in the leading order.
Let's take $Fig.1 (a)$ for instance, the spectator quark can be
attached to each of the quark coming from the 4-quark operators
with a hard gluon.

For the $B^0(\bar{B}^0) \to K^{\ast0} \bar{K}^{\ast0}$ decays
($Fig.1$), only the operators $O_{3-10}$ contribute via penguin
topology with light quark $q = s$ ($diagram\ a$) and via
annihilation topology with the light quark $q = d$ ($diagram\ b\
and\ c$) or $s$ ($diagram\ d$). It is a pure penguin mode with
only one kind of CKM element, as a result, it will not generate
any difference between $B^0$ and $\bar{B}^0$ decay and hence no
direct CP violation. Using the PQCD power counting rules
\cite{keum}, We can first predict that the main contribution came
from the factorizable parts of the emission diagram $F_{Le4}$(F
stands for factorizable, L stands for longitudinal, e stands for
emission and $4$ stands for the operator involved) with a large
Wilson coefficient $C_4+C_3/3-C_{10}/2-C_9/6$. The operator $O_6$
disappear here because the vector meson $\bar K^{*0}$ can not be
produced through a $(S-P)(S+P)$ operator. But in $B \to K^0 \bar
K^0$ decay \cite{kk}, there isn't such constraint and the
predicted branching ratio is about three times bigger than ours.
Second, the transverse parts of the emission diagram($F_{Ne4}$ and
$F_{Te4}$) are down by a factor $r_{k^*} (r_{K^*}\simeq
m_{K^*}/m_B)$ or $r_{K^*}^2$, then the longitudinal
parts($F_{Le4}$) dominate this process and give a large
longitudinal polarization fraction. Third, nonfactorizable
amplitudes $M$, including both emission and annihilation diagrams,
are suppressed by a power of $\bar \Lambda / M_B$ when compared
with factorizable ones. At last, we can forecast the factorizable
parts of the annihilation diagrams ($c,d$) counteract separately
in most of the cases, to be exactly, $F_{La3(5)}$ and $F_{Na(5)}$
vanish and $F_{Ta3(5)}$ survive but suppressed by $r_{k^*}^2$,
this makes the emission diagram relatively more important. The
factorizable parts for the space-like annihilation diagram ($b$)
with operator $(S-P)(S+P)$ and Wilson coefficient
$C_6+C_5/3-C_8/2-C_7/6$ do not counteract in any case but is still
not big enough to play the most important role. It is about 10
times smaller than the emission ones after calculation.
\begin{figure}[tbh]
\begin{center}
\epsfxsize=4.0in\leavevmode\epsfbox{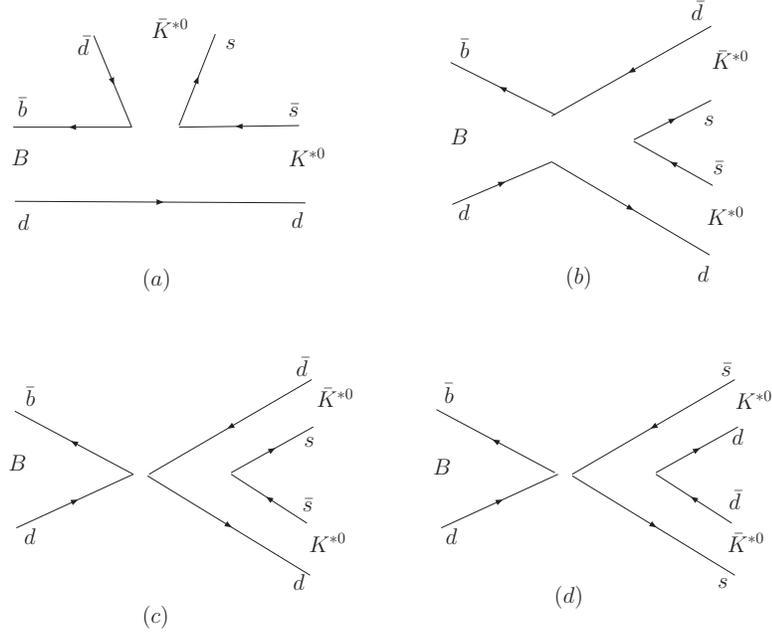}
\end{center}
\caption{{\protect\footnotesize Diagrams for $B^0\rightarrow
K^{*0} \bar K^{*0}$}}
\end{figure}

    Things are different for the $B^{\pm} \to K^{\ast\pm}
\bar{K}^{\ast0}(K^{\ast0})$ decays($Fig.2$). The operators
$O_{3-10}$ contribute via penguin topology with the light quark $q
= s$ ($diagram\ a$) and via the annihilation topology with $q = u$
($diagram\ b$), while tree operator $O_{1,2}$ also contribute via
annihilation topology ($diagram\ c$). We can see that there are
two kinds of CKM elements, $V_u$ from tree and $V_t$ from penguin,
that will induce weak phase and CP violation. We can get diagram
2.(a) and 2.(b) by replacing the d quark in $Fig.1$ (a) and (b) to
u quark. It makes no difference for power counting and the
conclusion we have given doesn't change. Diagram (c) is a tree
diagram, we have a much bigger Wilson coefficient $C_2+C_1/3$ for
the factorizable parts and $C_1$ for the nonfactorizable parts,
but at the same time, it is an annihilation diagram, as we have
stated, $F_L$ and $F_N$ vanish and $F_T$ is suppressed by
$r_{K^*}^2$ in this kind of diagram. After taken account of all
these two aspects, we can foresee that this diagram will be big
but not big enough to increase the branching ratio largely, we
also believe the transverse parts will play a more important role
than that of the former channel. Our calculation is consistent
with our predictions and will be shown in next section.

\begin{figure}[tbh]
\begin{center}
\epsfxsize=5.0in\leavevmode\epsfbox{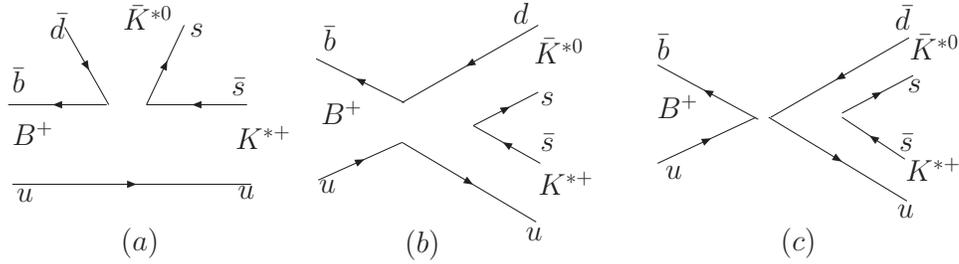}
\end{center}
\caption{{\protect\footnotesize Diagrams for $B^+\rightarrow
K^{*+} \bar K^{*0}$}}
\end{figure}

   We put the diagrams of $B^0(\bar{B}^0) \to K^{\ast+}K^{\ast-}$ decays in Fig.3.
From the topology we know that $O_{3-10}$ contribute via
annihilation topology with the light quark $q = u$ ($diagram\ b$)
or $s$ ($diagram\ c$) and tree operator $O_{1,2}$ contribute via
annihilation topology ($diagram\ a$). CPV occurs in this channels
for the same reason we have given for $B^+ \to K^{*+} K^{*0}$. But
when referring to the branching ratio, it is far different from
the former two, cause it's a pure annihilation mode and the
emission diagram which gives the main contribution to the
branching ratio of the former two channels no longer exist in this
process, so we can expect a smaller branching ratio for this
channel. Besides, the tree diagram involve $C_1+C_2/3$ for
factorizable parts and $C_2$ for nonfactorizble parts. As is
well-known, $C_1+C_2/3$ is small (about 0.1 when $t=4.8GeV$) but
$C_2$ is as big as 1.1 ($t=4.8GeV$), so the factorizable parts can
not be so important as the former two. Indeed, we found the
nonfactorizable tree diagram is the biggest one though it is
nonfactorizable suppressed after calculation. All our calculations
fit well with the predictions and they are shown in section 3.

\begin{figure}[tbh]
\begin{center}
\epsfxsize=5.0in\leavevmode\epsfbox{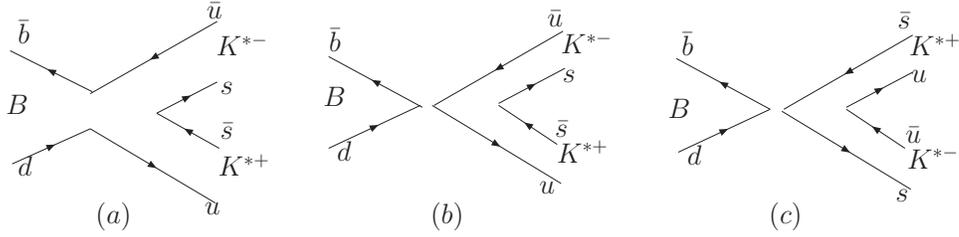}
\end{center}
\caption{{\protect\footnotesize Diagrams for $B^0\rightarrow
K^{*+}
 K^{*-}$}}
\end{figure}

  Now we are going to extract these decay channels within the PQCD
framework. For convenience, We adopt the light-cone coordinate
system \cite{lcsr}, then the four-momentum of the B meson and the
two $K^*$ mesons in the final state can be written as:
\begin{eqnarray}
\nonumber &&P_1=\frac{M_B}{\sqrt{2}}(1,1,{\bf0_T}),\\
\nonumber &&P_2=\frac{M_B}{\sqrt{2}}(1-r^2_{K^*},r^2_{K^*},{\bf 0_T}),\\
&&P_3=\frac{M_B}{\sqrt{2}}(r^2_{K^*},1-r^2_{K^*},{\bf0_T}),
\end{eqnarray}
in which $r_{K^*}$ is defined by
$r^2_{K^*}=\frac{1}{2}(1-\sqrt{1-4M_{K^*}^2/M_{B}^2})\simeq
M_{K^*}^2/M_{B}^2\ll 1$. To extract the helicity amplitudes, we
parameterize the polarization vectors. The longitudinal
polarization vector must satisfy the orthogonality and
normalization  : ${\epsilon_{2L} \cdot P_2}=0,\,\,{ \epsilon_{3L}
\cdot P_3 }=0 $, and ${ \epsilon_{2L}}^2={ \epsilon_{3L}}^2=-1$.
Then we can give the manifest form as follows:
\begin{eqnarray}
\nonumber{ \epsilon_{2L}}=\frac{1}{\sqrt{2}r_{K^*}}(1-r^2_{K^*},-r^2_{K^*},{\bf0_T}),\\
{
\epsilon_{3L}}=\frac{1}{\sqrt{2}r_{K^*}}(-r^2_{K^*},1-r^2_{K^*},{\bf0_T}).
\end{eqnarray}
As to the transverse polarization vectors, we can choose the
simple form:
\begin{eqnarray}
\nonumber{ \epsilon_{2T}}=\frac{1}{\sqrt{2}}(0,0,{\bf1_T}),\\
{ \epsilon_{3T}}=\frac{1}{\sqrt{2}}(0,0,{\bf1_T}).
\end{eqnarray}

The decay width  for these channels is :
\begin{eqnarray}
\Gamma=\frac{G_F^2{\bf |P_c|}}{16\pi M_B^2}\sum\limits_{\sigma
=L,T}{\cal M}^{\sigma\dag}{\cal M}^{\sigma}
\end{eqnarray}
where ${\bf P_c}$ is the 3-momentum of the final state meson, and
$|{\bf P_c}|=\frac{M_B}{2}(1-2r^2_{K^*})$. ${\cal M}^{\sigma}$ is
the decay amplitude which is decided by QCD dynamics, will be
calculated later in PQCD approach. The subscript $\sigma$ denotes
the helicity states of the two vector mesons with L(T) standing
for the longitudinal (transverse) components. After analyzing the
Lorentz structure, the amplitude can be decomposed into:
\begin{eqnarray}
{\cal M}^{\sigma}=M_B^2{\cal M}_L+M_B^2{\cal
M}_N\epsilon^{\ast}_2(\sigma=T)\cdot \epsilon^{\ast}_3(\sigma=T)+
i{\cal
M}_T\epsilon_{\mu\nu\rho\sigma}\epsilon_2^{\mu\ast}\epsilon_3^{\nu\ast}P_2^{\rho}P_3^{\sigma}.
\end{eqnarray}
We can define the longitudinal $H_0$, transverse $H_{\pm}$
helicity amplitudes
\begin{equation}
H_0=M^2_B{\cal M}_L,\,\,H_{\pm}=M^2_B{\cal M}_N\mp
M_{K^*}^2\sqrt{r^{\prime 2}-1}{\cal M}_T,
\end{equation}
where $r^{\prime}=\frac{P_2\cdot P_3}{M_{K^*}^2}$. After the
helicity summation, we can deduce that they satisfy the relation
\begin{equation}
\sum\limits_{\sigma =L,R}{\cal M}^{\sigma\dag}{\cal
M}^{\sigma}=|H_0|^2+|H_+|^2+|H_-|^2.
\end{equation}

There is  another equivalent set of definition of helicity
amplitudes
\begin{eqnarray}
\nonumber &&A_0=-\xi M^2_B{\cal M}_L,\\
\nonumber && A_{\|}=\xi \sqrt{2} M^2_B{\cal M}_N,\\
&& A_{\perp}=\xi M_{K^*}^2\sqrt{r^{\prime 2}-1}{\cal M}_T,
\end{eqnarray}
with $\xi$ the normalization factor to satisfy
\begin{eqnarray}
 |A_0|^2+|A_{\|}|^2+|A_{\perp}|^2=1,
\end{eqnarray}
 where the notations
$A_0$, $A_{\|}$, $A_{\perp}$ denote the longitudinal, parallel,
and perpendicular polarization amplitude.

What is followed is to calculate the matrix elements ${\cal M}_L$,
${\cal M}_N$ and ${\cal M}_T$ of the operators in the weak
Hamiltonian with PQCD approach. We have to admit the light cone
wave functions of mesons are not calculable in principal in PQCD,
but they are universal for all the decay channels. So that they
can be constraint from the measured other decay channels, like
$B\rightarrow K\pi$ and $B\rightarrow \pi\pi$ decays etc
\cite{keum,pipi}. For the heavy B meson, we have
\begin{eqnarray}
\frac{1}{\sqrt{2N_c}}(\not\!P_1+M_B)\gamma_5\phi_B(x,b)\label{bwave}.
\end{eqnarray}
For longitudinal polarized $K^*$ meson,
\begin{eqnarray}
\frac{1}{\sqrt{2N_c}}[M_{K^*}\not\!\epsilon_{2L}\phi_{K^*}(x)+\not\!\epsilon_{2L}
\not\!P_2\phi_{K^*}^t(x)+M_{\phi}I\phi^s_{K^*}(x)],
\end{eqnarray}
and for transverse polarized $K^*$ meson,
\begin{eqnarray}
\frac{1}{\sqrt{2N_c}}[M_{K^*}\not\!\epsilon_{2T}\phi^v_{K^*}(x)+\not\!\epsilon_{2T}
\not\!P_2\phi_{K^*}^T(x)+\frac{M_{K^*}}{P_2\cdot
n_-}i\epsilon_{\mu\nu\rho\sigma}\gamma_5\gamma^{\mu}\epsilon_{2T}^{\nu}P_2^{\rho}n_-^{\sigma}\phi^a_{K^*}(x)].
\end{eqnarray}
In the following concepts, we omit the subscript of the $K^*$
meson for simplicity.

The hard amplitudes are channel dependent, but they are
perturbative calculable. The amplitudes for $B^0 \to
K^{\ast0}\bar{K}^{\ast0}$ and $\bar{B}^0\to
K^{\ast0}\bar{K}^{\ast0}$are written as
\begin{eqnarray}
{\cal M}_{H}&=&f_{K^*}V_t^{*}F_{He4}+f_{B}V_t^{*}\left(\sum
_{i=3}^6F_{Hai}^{(d)}+\sum_{i=3,5}F_{Hai}^{(s)}\right)\nonumber \\
 && +V_t^*\left(\sum_{i=3,5}{\cal M}_{Hei}+\sum_{i=3}^6{\cal M}_{Hai}^{(d)}+\sum_{i=4,6}{\cal
 M}_{Hai}^{(s)}\right),
\\
\bar {\cal M}_{H}&=&f_{K^*}V_tF_{He4}+f_{B}V_t\left(\sum
_{i=3}^6F_{Hai}^{(d)}+\sum_{i=3,5}F_{Hai}^{(s)}\right)\nonumber \\
 && +V_t\left(\sum_{i=3,5}{\cal M}_{Hei}+\sum_{i=3}^6{\cal M}_{Hai}^{(d)}+\sum_{i=4,6}{\cal
 M}_{Hai}^{(s)}\right).
\end{eqnarray}
respectively, where the subscript $H=L,N,T$ denotes different
helicity amplitudes, and e(a) denotes the emission(annihilation)
topology. The hard parts for the factorizable amplitudes $F$ and
for the nonfactorizable amplitudes $\cal M$ are derived by
contracting the wave function to the lowest-order
one-gluon-exchange diagrams.

The helicity amplitudes ${\cal M}^+$ and ${\cal M}^-$
corresponding to $B^+ \to K^{*+}\bar K^{*0}$ and $B^- \to K^{*-}
K^{*0}$ are written as
\begin{eqnarray}
{\cal M}_{H}^+&=&V_u^*T_H-V_t^{*}P_H\label{m+}
\\
{\cal M}_{H}^-&=&V_uT_H-V_tP_H\label{m-}
\end{eqnarray}with \begin{eqnarray}T_H=f_{B}F_{Ha2}^{(u)}+{\cal
M}_{Ha1}^{(u)}\nonumber\end{eqnarray} and
\begin{eqnarray}
P_H=f_{K^*}F_{He4}+f_{B}\sum_{i=4,6}F_{Hai}^{(d)}
+\sum_{i=3,5}{\cal M}_{Hei}+\sum_{i=3,5}{\cal
M}_{Hai}^{(d)}\nonumber.
\end{eqnarray}

The helicity amplitudes for $B^0 \to K^{*+} K^{*-}$ and $\bar B^0
\to K^{*+} K^{*-}$ are written as
\begin{eqnarray}
{\cal M}_{H}^{\prime}&=&V_u^*T_H^{\prime}-V_t^{*}P_H^{\prime}\nonumber\\
&=&V_u^*\left(f_BF_{Ha1}^{(u)}+{\cal
M}_{Ha2}^{(u)}\right)-V_t^{*}\sum_{q=u,s}\left(f_{B}\sum_{i=3,5}F_{Hai}^{(q)}
+\sum_{i=4,6}{\cal M}_{Hai}^{(q)}\right),\label{mprime}
\\
\bar {\cal M}_{H}^{\prime}&=&V_uT_H^{\prime}-V_tP_H^{\prime}\nonumber\\
&=&V_u\left(f_BF_{Ha1}^{(u)}+{\cal
M}_{Ha2}^{(u)}\right)-V_t\sum_{q=u,s}\left(f_{B}\sum_{i=3,5}F_{Hai}^{(q)}
+\sum_{i=4,6}{\cal M}_{Hai}^{(q)}\right),\label{barmprime}
\end{eqnarray}
where the $F_{Lai}^{(q)}$ and $F_{Nai}^{(q)}$ vanish and
$F_{Tai}^{(q)}$ takes from of Eqs.(\ref{fta3},\ref{fta5}). The
detailed formulas with polarization ${\cal M}_L$, ${\cal M}_N$,
and ${\cal M}_T$ for each diagram are given in the appendix.

\section{Numerical analysis}
\hspace*{\parindent} For the B meson wave function used in
eq.(\ref{bwave}), we employ the model \cite{keum,pipi,bwave}
\begin{eqnarray}
\phi_B(x)=N_Bx^2(1-x)^2\exp\left[-\frac{1}{2}(\frac{xM_B}{\omega_B})^2-\frac{\omega^2_Bb^2}{2}\right],
\end{eqnarray}
where the shape parameter $\omega_B=0.4GeV$ has been constrained
in other decay modes. The normalization constant $N_B=91.784GeV$
is related to the B decay constant $f_B=0.19GeV$. It is one of the
two leading twist B meson wave functions; the other one is power
suppressed, so we omit its contribution in the leading power
analysis \cite{cdluatepjc28}. The $K^*$ meson distribution
amplitude up to twist-3 are given by \cite{bbkt} with QCD sum
rules.
\begin{eqnarray}
&&\phi_{K^*}(x)=\frac{3f_{K^*}}{\sqrt{2N_c}}x(1-x)[1+0.57(1-2x)+0.07C_2^{3/2}(1-2x)],\\
&&\phi^t_{K^*}(x)=\frac{f_{K^*}^T}{2\sqrt{2N_c}}\left\{0.3(1-2x)[3(1-2x)^2+10(1-2x)-1]
+1.68C_4^{1/2}(1-2x)\right.\nonumber\\
&&
\left.\ \ \ \ \ \ \ \ \ \ \ \ +0.06(1-2x)^2[5(1-2x)^2-3]+0.36\left\{1-2(1-2x)[1+\ln(1-x)]\right\}\right\},\\
&&\phi^s_{K^*}(x)=\frac{f_{K^*}^T}{2\sqrt{2N_c}}\left\{3(1-2x)\left[1+0.2(1-2x)+0.6(10x^2-10x+1)\right]
\right.\nonumber\\
&& \left.\ \ \ \ \ \ \ \ \ \ \ \
-0.12x(1-x)+0.36[1-6x-2\ln{(1-x)}]\right\}.
\end{eqnarray}
\begin{eqnarray}
&&\phi^T_{K^*}(x)=\frac{3f_{K^*}^T}{\sqrt{2N_c}}x(1-x)[1+0.6(1-2x)+0.04C_2^{3/2}(1-2x)
],\\
&&\phi^v_{K^*}(x)=\frac{f_{K^*}^T}{2\sqrt{2N_c}}\left\{\frac{3}{4}[1+(1-2x)^2+0.44(1-2x)^3]
\right.\nonumber\\
&& \left.\ \ \ \ \ \ \ \ \ \ \ \ +0.4C_2^{1/2}(1-2x)
+0.88C_4^{1/2}(1-2x)+0.48[2x+\ln(1-x)]
\right\},\\
&&\phi^a_{K^*}(x)=\frac{f_{K^*}^T}{4\sqrt{2N_c}}\left\{3(1-2x)[1+0.19(1-2x)+0.81(10x^2-10x+1)]
\right.\nonumber\\
&& \left.\ \ \ \ \ \ \ \ \ \ \ \
-1.14x(1-x)+0.48[1-6x-2\ln(1-x)]\right\},
\end{eqnarray}
where the Gegenbauer polynomials are
\begin{eqnarray}
&&C_2^{\frac{1}{2}}(\xi)=\frac{1}{2}(3\xi^2-1),\\
&&C_4^{\frac{1}{2}}(\xi)=\frac{1}{8}(35\xi^4-30\xi^2+3),\\
&&C_2^{\frac{3}{2}}(\xi)=\frac{3}{2}(5\xi^2-1).
\end{eqnarray}

In paper \cite{resolveploar}, Li has suggested to reanalyze the
$K^*$ meson distribution amplitude in order to solve the
polarization puzzle of $B \to \phi K^*$. In that channel, Babar
\cite{babarphikstar} and Belle \cite{bellephikstar} have reported
a longitudinal polarization fraction($R_L$) small to $50 \%$, it
is different from most theoretical predictions and is considered
as a puzzle. Many discussions have been given
\cite{otherresolve1,otherresolve2,otherresolve3,otherresolve4} and
among which Hsiang-nan Li argued a smaller $B \to K^*$ form
factor($A_0 \approx 0.3$), which doesn't contradict any existing
data, and hence a new distribution amplitude for $K^*$ meson. Any
how, this assumption need to be justified by experiment and we
will take the traditional wave function in this letter. If future
experiment confirms a smaller $B \to K^{*}$ form factor and those
argues, we just replace the wave function and get a smaller $R_L$
(about $65 \%$) and smaller branching ratios (about
$3\times10^{-7}$) immediately. On the other hand, if future
experiments find a small $R_L$ and branching ratios for $B^0(B^+)
\to K^{*0}(K^{*+})\bar K^{*0}$, it may be a support for a smaller
$B \to K^*$ form factor and the validity of PQCD.

 We employ the constants as follows \cite{particle}: the Fermi coupling constant
$G_F=1.16639\times10^{-5}GeV^{-2}$, the meson masses
$M_B=5.28GeV,\ M_{K^*}=0.89GeV$, the decay constants
$f_{K^*}=0.217GeV,\ f_{K^*}^T=0.16GeV$ \cite{fkstar}, the central
value of the CKM matrix elements $|V_{td}|=0.0075,\
|V_{tb}|=0.9992,\ |V_{ud}|=0.9745,\ |V_{ub}|=0.0033$ and the B
meson lifetime $\tau_{B^0}=1.536ps$$\ (\tau_{B^\pm}=1.671ps)$
\cite{particle}.

   If we choose the CKM phase angle $\alpha(\varphi_2)=95^{\circ}$ \cite{particle}, then our
our numerical results are given in TABLE.1, where
$\phi_{\parallel}\equiv Arg(A_{\parallel}/A_0)$ and
$\phi_{\perp}\equiv Arg(A_{\perp}/A_0)$. From the table we are
convinced more with our power counting stated in chapter 2. We
also find the polarization fraction $R_{\|}\simeq R_{\perp}$ and
relative phase is around 2.6 for the former 3 channels. This is
good news both for us and PQCD, since the current $B \to \phi K^*$
data \cite{bellephikstar,babarphikstar}, which is also governed by
the $B\to K^*$ form factors, suggest $R_{\|}\simeq R_{\perp}$,
$\phi_{\|}\simeq 2.3$ and $\phi_{\perp}\simeq 2.5$. These data are
contrary from those rescattering effects \cite{otherresolve4} and
seem to support the evaluation of the relative strong phase in
PQCD.

\[
\begin{tabular}{ccccccc}
\multicolumn{7}{c}{TABLE.{1}.\ \ Helicity amplitudes and relative
phases} \\ \hline \hline
Channel & BR$(10^{-7})$ & $|A_0|^2$ & $|A_{\|}|^2$ & $|A_{\perp}|^2$ & $\phi_{\|}(rad)$ & $\phi_{\perp}(rad)$ \\
\hline \hline $B^0(\bar{B}^0) \to K^{\ast0}\bar{K}^{\ast0}$ & 3.5
& 0.78 & 0.12 & 0.10 & 2.8 & 2.8\\
$B^+ \to K^{*+}\bar K^{*0}$&4.0&0.75&0.10&0.15&2.6&2.4\\
$B^- \to K^{*-} K^{*0}$&5.5&0.88&0.08&0.04&2.7&3.0\\
$B^0 \to K^{*+} K^{*-}$&0.22&0.99&0.005&0.005&4.1&2.2\\
$\bar B^0 \to K^{*+} K^{*-}$&1.1&0.99&0.005&0.003&3.6&1.9\\
\hline\hline
\end{tabular}\]

To test the contribution from different parts separately, we take
$B^0(\bar{B}^0) \to K^{\ast0}\bar{K}^{\ast0}$ for example and
classify the contributions into 4 kinds (see TABLE.2.): (1) full
contribution, (2) without annihilation nor nonfactorizable
contributions, (3) without annihilation contributions, (4) without
nonfactorizable contributions. Form the table we are convinced the
annihilation contribution play an important role to the branching
ratio. The annihilation diagrams counteract with emission diagram
severely, it even makes the branching ratio smaller when compared
with the pure emission contribution. We also notice that
contribution from the factorizable parts of the emission diagram
is also bigger than the total branching ratio for $B(\bar{B}) \to
K^{\ast0}\bar{K}^{\ast0}$. As a result, if we change the form
factor $A_0$ from 0.4 to 0.32 as \cite{resolveploar}, then our
calculation give a branching ratio of $2.3\times 10^{-7}$.

\begin{center}
\begin{tabular}{ccccccc} \multicolumn{7}{c}{TABLE.{2}.\ \
Contribution from different parts: }\\
 \multicolumn{7}{c}{(1) full contribution,
(2) without annihilation }\\
\multicolumn{7}{c}{nor nonfactorizable contributions, (3) without
}\\
\multicolumn{7}{c}{annihilation contributions, (4) without nonfa}\\
\multicolumn{7}{c}{-ctorizable
contributions.\ \ \ \ \ \ \ \ \ \ \ \ \ \ \ \ \ \ \ \ \ \ \ \ \ \ \ \ }\\
 \hline \hline
Class & BR$(10^{-7})$ & $|A_0|^2$ & $|A_{\|}|^2$ & $|A_{\perp}|^2$ & $\phi_{\|}(rad)$ & $\phi_{\perp}(rad)$ \\
\hline \hline
(1)&3.5&0.78&0.12&0.10&2.8&2.8\\
(2)&4.4&0.94&0.03&0.03&$\pi$&$\pi$\\
(3)&3.8&0.86&0.07&0.07&3.3&3.3\\
(4)&4.1&0.87&0.07&0.07&2.5&2.6\\
 \hline\hline
\end{tabular}
\end{center}

For $B^+ \to K^{*+}\bar K^{*0}$ and $B^- \to K^{*-} K^{*0}$, it is
similar to do so, we find annihilation diagrams contribute $7\%$
and $31\%$ to the total branching ratio respectively. If we take a
smaller form factor and immediately get these values shown in
TABLE.3. We have to say these results are rather roughly because
no precision wave function for $K^*$ have been given in that
paper.

\begin{center}
\begin{tabular}{ccccc}
\multicolumn{5}{c}{TABLE.{3}.\ \ The impact of a smaller form }\\
\multicolumn{5}{c}{factor $A_0=0.32$ on different
channels}\\\hline \hline
Decay Channel & BR$(10^{-7})$&$\ |A_0|^2\ $&$\ |A_{\|}|^2\ $&$|A_{\perp}|^2$\\
\hline \hline $B^0(\bar{B}^0) \to K^{\ast0}\bar{K}^{\ast0}$& 2.3
& 0.67 &0.18&0.15\\
$B^\pm \to K^{*\pm}\bar K^{*0}(K^{*0})$&3.3& 0.75& 0.13&0.12\\
 \hline\hline
\end{tabular}
\end{center}

    To extract the CPV parameter of $B^+ \to K^{*+}\bar K^{*0}$ and
$B^- \to K^{*-} K^{*0}$, we can rewrite the helicity amplitude in
(\ref{m+},\ref{m-}) as a function of the CKM phase angle $\alpha$:
\begin{eqnarray}
{\cal M}_{H}^+&=&V_u^*T_H-V_t^{*}P_H \nonumber\\
&=&V_u^*T_H(1+Z_He^{i(\alpha+{\delta}_H)})\label{m++}\\
 {\cal
M}_{H}^-&=&V_uT_H-V_tP_H\nonumber\\
&=&V_uT_H(1+Z_He^{i(-\alpha+{\delta}_H)})\label{m--}
\end{eqnarray}where $Z_H=|V_t^*/V_u^*||P_H/T_H|$, and $\delta$ is the
relative strong phase between tree($T$) and penguin($P$) diagrams.
Here in PQCD approach, the strong phase comes from the
nonfactorizable diagrams and annihilation diagrams. This can be
seen from Eqs.(\ref{ha},\ref{hd},\ref{hf}), where the modified
Bessel function has an imaginary part. This is different from FA
\cite{bsw} and Beneke-Buchalla-Neubert-Sachrajda(BBNS) \cite{bbns}
approaches. In that approaches, annihilation diagrams are not
taken into account, strong phases mainly come from the so-called
Bander-Silverman-Soni mechanism \cite{bss}. As shown in
\cite{keum}, these effects are in fact
next-to-leading-order($\alpha_s$ suppressed) elements and can be
neglect in PQCD approach. We give the averaged branching ratio of
$B^\pm \to K^{*\pm}\bar K^{*0}(K^{*0})$ as a function of $\alpha$
in Fig.4.

\begin{figure}[tbh]
\begin{center}
\epsfxsize=5.0in\leavevmode\epsfbox{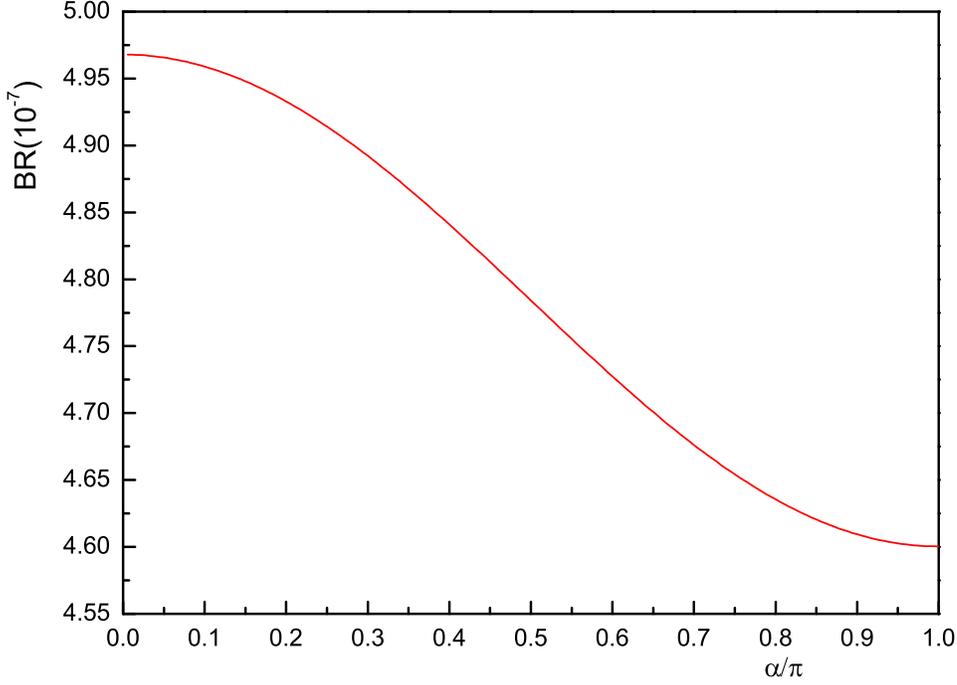}
\end{center}
\caption{{\protect\footnotesize Average branching ratios of
$B^{\pm}\rightarrow K^{*\pm}
 \bar K^{*0}(K^{*0})$ as a function of $\alpha$.
}}
\end{figure}

Using Eqs.(\ref{m++},\ref{m--}), the direct CP violating parameter
is
\begin{eqnarray}
A^{dir}_{CP}&=&\frac{{|M^+|}^2-{|M^-|}^2}{{|M^+|}^2+{|M^-|}^2}\nonumber
\\
&=&\frac{-2sin\alpha\left(T_L^2sin\delta_L+2T_N^2sin\delta_N+2T_T^2sin\delta_T
\right)}{T_L^2(1+Z_L^2+2Z_Lcos\alpha
cos\delta_L)+2\sum_{i=N,T}T_i^2\left(1+Z_i^2+2Z_icos\alpha
cos\delta_i\right) }.\label{adircp}
\end{eqnarray}
Since the transverse polarization is twice of freedom when
comparing with longitudinal one, the factor before $T_N$ and $T_T$
is twice as $T_L$. If we choose $\alpha$ as $95^\circ$, then the
direct CP asymmetry $A_{CP}^{dir}$ for these channels are:
\begin{eqnarray}
A_{CP}^{dir}(B^0(\bar B^0)\to K^{*0}\bar K^{*0})&=&0,\\
 A_{CP}^{dir}(B^{\pm} \to K^{\ast\pm}
\bar{K}^{\ast0}(K^{\ast0}))&=&-15\%,\\
A_{CP}^{dir}(B^0(\bar B^0)\to K^{*+}K^{*-})&=&-65\%.
\end{eqnarray}
We notice the CP asymmetry of $B^0(\bar B^0)\to K^{*0}\bar K^{*0}$
is zero, since only pure penguin contribution in this channel. The
CP asymmetry of $B^{\pm} \to K^{\ast\pm}
\bar{K}^{\ast0}(K^{\ast0})$ is relatively small but large in
$B^0(\bar B^0)\to K^{*+}K^{*-}$, this is consistent with PQCD
prediction. Using the power counting rules we stated in section.2,
the former channel is penguin dominated while the latter one is
tree dominated, then from the definition of $Z_H$ we can easily
deduce a big $Z_H$ for the former channel and a small $Z_H$ for
the latter one, so we can forecast a similar conclusion using
Eqs.(\ref{adircp}) without any calculation. We also notice the CP
asymmetry for these channels are sensitive to $\alpha$, hence we
put $A_{CP}^{dir}$ as a function of $\alpha$ in Fig.5.

\begin{figure}[tbh]
\begin{center}
\epsfxsize=5.0in\leavevmode\epsfbox{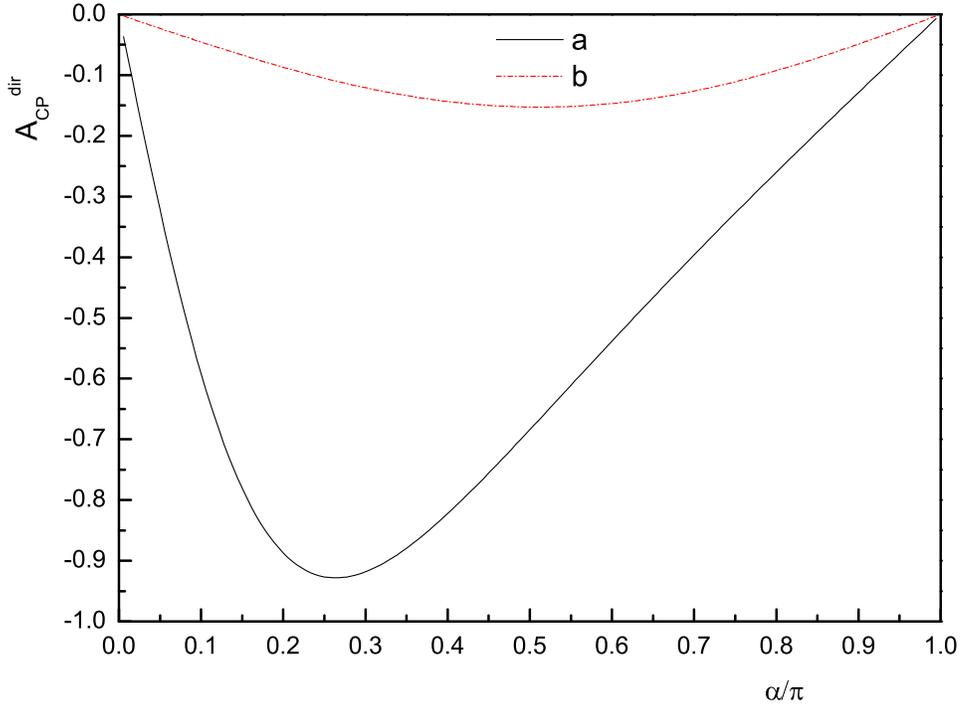}
\end{center}
\caption{{\protect\footnotesize $A^{dir}_{CP}$ as a function of
$\alpha$. (a) $B^0(\bar B^0) \to K^{*+}K^{*-}$, (b) $B^\pm \to
K^{*\pm}\bar K^{*0}$.}}
\end{figure}

For the $B^0(\bar B^0)\to K^{*+}K^{*-}$ decays, it is hard to
distinguish $B^0$ and $\bar B^0$, we can use the value given in
TABLE.1 to get an average branching ratio of $6.3\times 10^{-8}$.
If we let CKM angle $\alpha$ as a free parameter, then the
evaluation of averaged branching ratio is shown in Fig.6.

\begin{figure}[tbh]
\begin{center}
\epsfxsize=5.0in\leavevmode\epsfbox{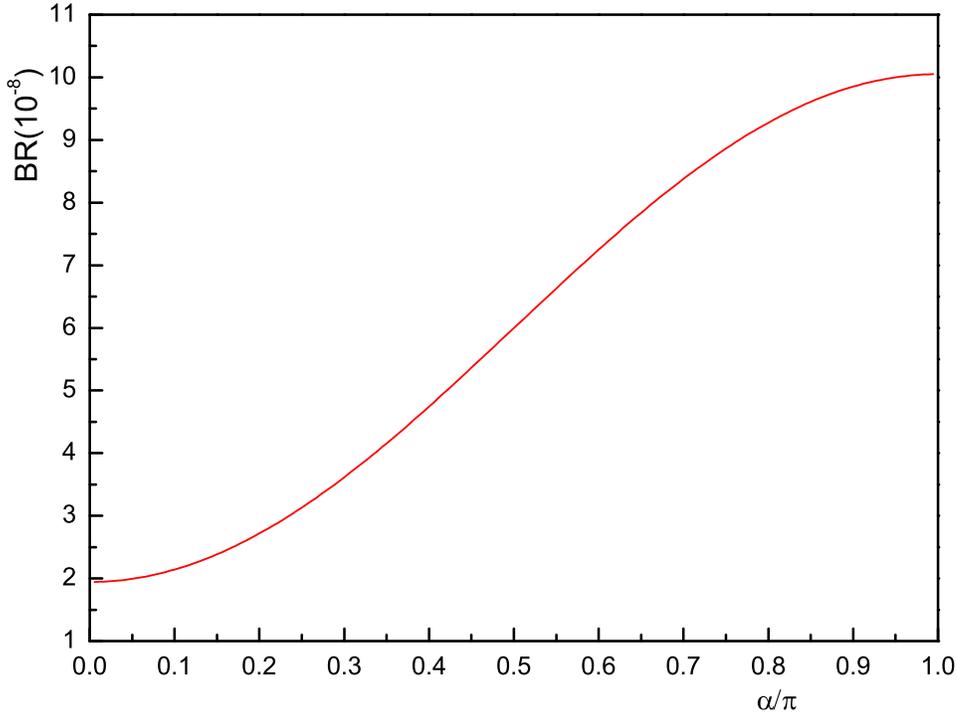}
\end{center}
\caption{{\protect\footnotesize Average branching ratios of
$B^0(\bar B^0) \to K^{*+}K^{*-}$ as a function of $\alpha$.}}
\end{figure}

When it refered to CP asymmetry of $B^0(\bar B^0)$ decays, it is
more complicated due to the $B^0-\bar B^0$ mixing. The CP
asymmetry is time dependent:
\begin{eqnarray}
A_{CP}(t)\simeq A^{dir}_{CP}cos(\Delta
mt)+a_{\epsilon+\epsilon^{\prime}}sin(\Delta mt)\label{acpt}
\end{eqnarray}
where $\Delta m$ is the mass difference of the two mass
eigenstates of neutral $B$ mesons. The direct CP violation
parameter $A^{dir}_{CP}$ is already defined in Eq.(\ref{adircp}),
while the mixing-related CP violation parameter is defined as
\begin{eqnarray}
a_{\epsilon+\epsilon^{\prime}}=\frac{-2Im(\lambda_{CP})}{1+|\lambda_{CP}|^2},
\end{eqnarray}
where
\begin{eqnarray}
\lambda_{CP}=\frac{V_t^*<f|H_{eff}|\bar
B|>}{V_t<f|H_{eff}|B>}.\label{lambdacp}
\end{eqnarray}
In these two channels, over $90\%$ of the branching ratios are
composed of longitudinal fraction, so we can neglect the
transverse contribution and use Eqs(\ref{mprime},\ref{barmprime})
to derive
\begin{eqnarray}
\lambda_{CP}\simeq
e^{2i\alpha}\frac{1+Z_Le^{i(\delta_L-\alpha)}}{1+Z_Le^{i(\delta_L+\alpha)}}.
\end{eqnarray}
People usually believe the penguin diagram contribution have been
suppressed when comparing with the tree contribution. i.e. in
$B(\bar B)\to \pi^+\pi^-$ decay, Wilson coefficients of penguin
diagram are loop suppressed, people believe $Z\ll1$,
$\lambda_{CP}\simeq exp[2i\alpha]$,
$a_{\epsilon+\epsilon^{\prime}}=-sin2\alpha$, and $
A^{dir}_{CP}\simeq0$, then it is easy to extract $sin2\alpha$
through the measurement of CPV. However, $Z$ is not very small and
$a_{\epsilon+\epsilon^{\prime}}$ is not a simple function of
$-sin2\alpha$ even in $B(\bar B)\to \pi^+\pi^-$ \cite{pipi}, hence
we couldn't get an exact $\alpha$ and this is called penguin
pollution. In our channel, it is similar. After taking into
account of the Wilson coefficient and the penguin enhancement we
have stated, we get a $Z_L$ as large as $0.80$ and
$\delta_L=2.35$. We put $a_{\epsilon+\epsilon^{\prime}}$ as a
function of $\alpha$ in Fig.7 and we can see there isn't a simple
relationship between $a_{\epsilon+\epsilon^{\prime}}$ and
$-sin2\alpha$.

\begin{figure}[tbh]
\begin{center}
\epsfxsize=5.0in\leavevmode\epsfbox{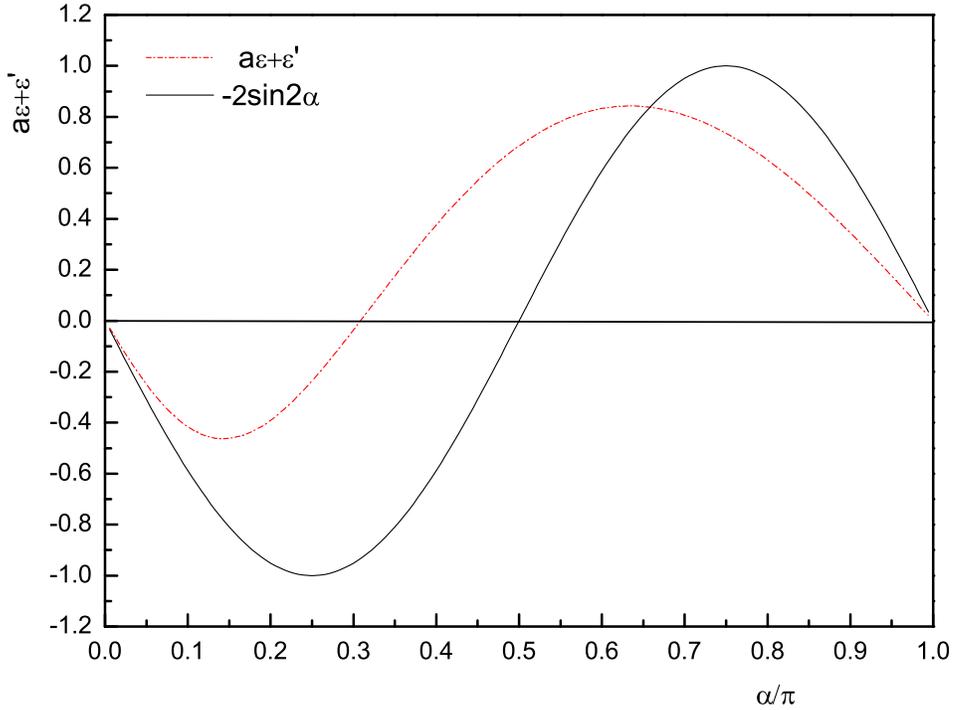}
\end{center}
\caption{{\protect\footnotesize CP violation parameters
$a_{\epsilon+\epsilon^{\prime}}$ of $B^0(\bar B^0) \to
K^{*+}K^{*-}$ as a function of $\alpha$}}
\end{figure}

If we integrate the time variable t of Eq.(\ref{acpt}), we will
get the total CP asymmetry as
\begin{eqnarray}
A_{CP}=\frac{1}{1+x^2}A_{CP}^{dir}+\frac{x}{1+x^2}a_{\epsilon+\epsilon^{\prime}}
\end{eqnarray}
with $x=\Delta m/\Gamma\simeq0.771$ for the $B^0-\bar B^0$ mixing
in SM \cite{particle}. The integrated CP asymmetries for $B^0(\bar
B^0) \to K^{*+}K^{*-}$ are shown in Fig.8. As for $B^0(\bar
B^0)\to K^{*0} \bar K^{*0}$, there is only penguin contribution in
this decay, direct CP is zero in Eqs.(\ref{adircp}). The weak
phase of penguin $V_{tb}V_{td}^*$ is cancelled by the $B^0-\bar
B^0$ mixing phase $V_{tb}^*V_{td}$, so $\lambda_{CP}$ (see
Eqs.(\ref{lambdacp})) is real here and
$a_{\epsilon+\epsilon^{\prime}}=0$. In fact, to the next leading
order, there is a small up quark and charm quark penguin
contribution which may give a small direct and mixing CP.

\begin{figure}[tbh]
\begin{center}
\epsfxsize=5.0in\leavevmode\epsfbox{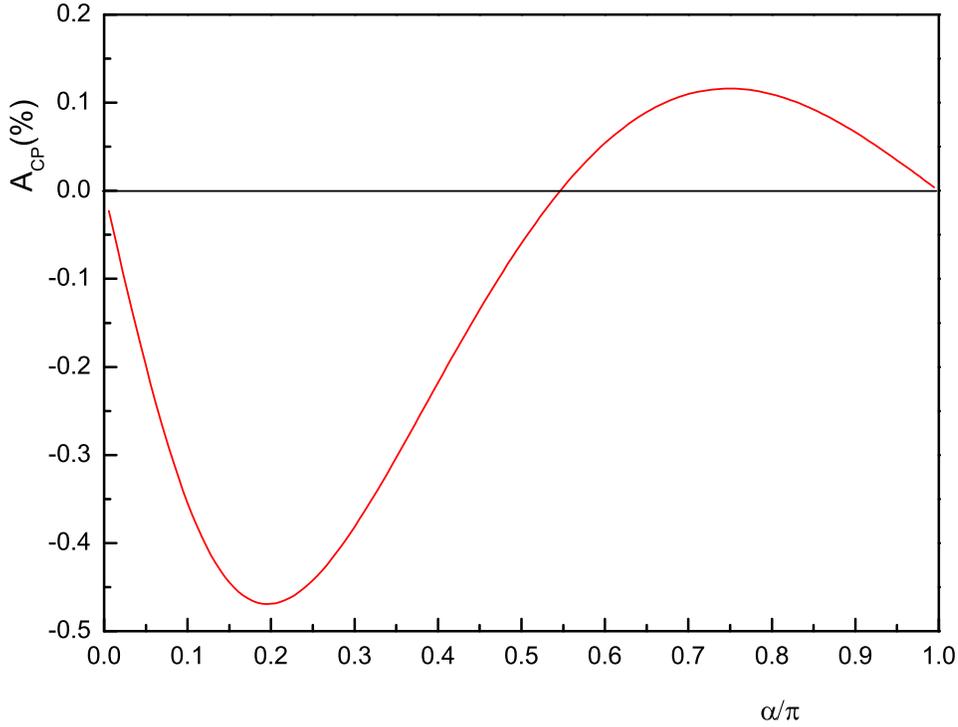}
\end{center}
\caption{{\protect\footnotesize $A_{CP}$ of $B(\bar B) \to
K^{*+}K^{*-}$ as a function of $\alpha$}}
\end{figure}

When the PQCD formalism is extended to $O(\alpha_s^2)$, the hard
scales can be determined more precisely and the scale independence
of our predictions will be improved. Before this calculation is
carried out, we consider the hard scales t located between
$0.75-1.25$ times the invariant masses of the internal particles.
For example, we take $t_e$(see Eqs.(\ref{tfact})) in the following
range,
\begin{eqnarray}
{\rm max}(0.75\sqrt{x_{2}}M_{B},1/b_{1},1/b_{3})&<t_{e}^{(1)}<&
{\rm max}(1.25\sqrt{x_{2}}M_{B},1/b_{1},1/b_{3})\;,
\nonumber \\
{\rm max}(0.75\sqrt{x_{1}}M_{B},1/b_{1},1/b_{3})&< t_{e}^{(2)}
<&{\rm max}(1.25\sqrt{x_{1}}M_{B},1/b_{1},1/b_{3}),
\end{eqnarray}
in order to estimate the $O(\alpha_s^2)$ corrections. Then we can
obtain the value area of the branching ratio for the penguin
dominated modes
\begin{eqnarray}
BR(B^0(\bar{B}^0) \to
K^{\ast0}\bar{K}^{\ast0})=(3.5^{+1.3}_{-0.7})\times 10^{-7},\\
BR(B^\pm \to K^{*\pm}\bar
K^{*0}(K^{*0}))=(4.8^{+1.2}_{-0.8})\times 10^{-7},
\end{eqnarray}
which is sensitive to the change of $t$, so we can estimate that
the next to leading order corrections will give about $20\%$
contribution. The ratios $R_0,\,\,R_{\|}\,\,$and $R_{\perp}$ are
not very sensitive to the variation of $t$, since the main
contribution $F_{Le4}$, $F_{Ne4}$ and $F_{Te4}$ vary similarly
when we conduct such changes on the maximum energy scale $t$. For
the tree dominated modes $B^0(\bar B^0) \to K^{*+}K^{*-}$, the
evaluation of the Wilson coefficients $C_1$ and $C_2$ is slow,
hence it is not sensitive to the scale $t$, we changed the
parameter $\omega_b$ of the B meson wave function from 0.32 to
0.48 and found the absolute value for each integration become
larger when $\omega_b=0.32$ and smaller when $\omega_b=0.48$, then
the value area for these decays are
\begin{eqnarray}
BR(B^0(\bar B^0) \to
K^{\ast+}{K}^{\ast-})=(6.4^{+0.5}_{-1.0})\times 10^{-8}.
\end{eqnarray}

If we compare our predictions with generalized FA \cite{ali}
\begin{eqnarray}
BR(B^0(\bar{B}^0) \to
K^{\ast0}\bar{K}^{\ast0})=3.5\times 10^{-7},\\
BR(B^\pm \to K^{*\pm}\bar K^{*0}( K^{*0}))=3.7\times 10^{-7},
\end{eqnarray}
with $N_c=3$ and QCDF \cite{chenghy}
\begin{eqnarray}
BR(B^0 \to
K^{\ast0}\bar{K}^{\ast0})=3.2\times 10^{-7},\\
BR(B^- \to K^{*-} K^{*0})=3.4\times 10^{-7},
\end{eqnarray}
with the form factor from Light cone sum rules(LCSR)
\cite{lcsr,lcsr2}, we can easily find out their predictions for
$B^\pm \to K^{*\pm}\bar K^{*0}(K^{*0})$ are a little smaller than
our's, and neither of them gives the branching ratios and CPV
parameters of $B^0(\bar B^0) \to K^{*+}\bar K^{*-}$, because
annihilation diagrams can not be calculated in FA or QCDF in
principle, while in PQCD all diagrams are calculated strictly. If
we drop the annihilation contributions, we can get a similar
results ($3.7\times10^{-7}$) with them and they are already shown
in TABLE.2. Current experiments \cite{particle} only give the
upper limit for these decays
\begin{eqnarray}
BR(B^0 \to
K^{\ast0}\bar{K}^{\ast0})<2.2\times 10^{-5},\\
BR(B^+ \to K^{*+} K^{*0})<7.1\times 10^{-5},\\
BR(B \to K^{\ast+}\bar{K}^{\ast-})<1.41\times 10^{-4},
\end{eqnarray}
and future experiments are expected.

\section{Summary}
\hspace*{\parindent}In this paper we have predicted the branching
ratios, polarization fraction and CP asymmetries of $B \to K^*K^*$
modes using PQCD theorem in SM. We perform all leading diagrams,
including both emission and annihilation diagrams, with up to
twist-3 wave functions. The predicted branching ratios are
compared with experiments and values from other approaches. We
analyze the contribution from each parts for each decay channel,
and found the annihilation diagrams is not very small to be
neglected, then we present the dependence of the CP asymmetry and
branching ratios on the CKM angle $\alpha$. We also discussed the
potential impact of a smaller form factor $A_0$ in our paper.

\section*{Acknowledgments}
\hspace*{\parindent}This work is partly supported by the National
Science Foundation of China under Grant (No.90103013, 10475085 and
10135060), We thank J-F Cheng, H-n. li, Y. Li, and X-Q Yu for
helpful discussions. We also thank F-Q Wu for solving the problems
in our programmes.

\begin{appendix}
\section{Factorization formulas}
The factorizable amplitudes are written as
\begin{eqnarray}
F_{Le4}&=&8\pi
C_{F}M_{B}^{2}\int_{0}^{1}dx_{1}dx_{2}\int_{0}^{\infty}
b_{1}db_{1}b_{2}db_{2}\Phi _{B}( x_{1},b_{1}) \ \left\{ \left
[(1+x_{2})\Phi_{K^*}(x_2)
\right.\right. \nonumber\\
&&+\left.
r_{K^*}(1-2x_{2})(\Phi^{t}_{K^*}(x_3)+\Phi^{s}_{K^*}(x_2))\right]
E_{e4}(t^{(1)}_{e}) h_{e}(x_{1},x_{2},b_{1},b_{2})
\nonumber\\
&&\left.+\left[ 2r_{K^*}\Phi^{s}_{K^*}(x_2)+r_{K^*}^2
\phi_{K^*}(x_2)\right] E_{e4}(t^{(2)}_{e})
h_{e}(x_{2},x_{1},b_{2},b_{1}) \right\}\;,
\\
F_{Ne4} &=&8\pi
C_{F}M_{B}^{2}\int_{0}^{1}dx_{1}dx_{2}\int_{0}^{\infty}
b_{1}db_{1}b_{2}db_{2}\Phi _{B}(x_{1},b_{1})
\nonumber\\
&& \times r_{K^*} \left\{ [\Phi^{T}_{K^*}(x_2)
+2r_{K^*}\Phi^{v}_{K^*}(x_2)+r_{K^*}x_{2}
(\Phi^{v}_{K^*}(x_2)-\Phi^{a}_{K^*}(x_2))]\right.
\nonumber \\
&& \times E_{e4}(t^{(1)}_{e}) h_{e}(x_{1},x_{2},b_{1},b_{2})
\nonumber \\
&&\left.+r_{K^*}[\Phi^{v}_{K^*}(x_2)+\Phi^{a}_{K^*}(x_2)]
E_{e4}(t^{(2)}_{e}) h_{e}( x_{2},x_{1},b_{2},b_{1})\right\}\;,
\\
F_{Te4} &=&16\pi
C_{F}M_{B}^{2}\int_{0}^{1}dx_{1}dx_{2}\int_{0}^{\infty}
b_{1}db_{1}b_{2}db_{2}\Phi _{B}(x_{1},b_{1})
\nonumber\\
&& \times r_{K^*} \left\{ [\Phi^{T}_{K^*}(x_2)
+2r_{K^*}\Phi^{a}_{K^*}(x_2)+r_{K^*}x_{2}
(\Phi^{a}_{K^*}(x_2)-\Phi^{v}_{K^*}(x_2))]\right.
\nonumber \\
&& \times E_{e4}(t^{(1)}_{e}) h_{e}(x_{1},x_{2},b_{1},b_{2})
\nonumber \\
&&\left.+r_{K^*}[\Phi^{v}_{K^*}(x_2)+\Phi^{a}_{K^*}(x_2)]
E_{e4}(t^{(2)}_{e}) h_{e}( x_{2},x_{1},b_{2},b_{1})\right\}\;,\\
F_{Ta1(2)}^{\left( u\right) } &=&32\pi
C_{F}M_{B}^{2}r_{k^*}^2\int_{0}^{1}dx_{2}dx_{3}\int_{0}^{\infty}
b_{2}db_{2}b_{3}db_{3}
\nonumber \\
&&\times
\left[(2-x_2)(\Phi^{v}_{k^*}(x_2)\Phi^{a}_{k^*}(x_3)-\Phi^{a}_{k^*}(x_2)\Phi^{v}_{k^*}(x_3))
\right.
\nonumber \\
&&\left.
+x_2(\Phi^{a}_{k^*}(x_2)\Phi^{a}_{k^*}(x_3)-\Phi^{v}_{k^*}(x_2)\Phi^{v}_{k^*}(x_3))\right]
\nonumber\\
&&\times
E_{a1(2)}^{(u)}(t_{au}^{(1)})h_{a}(1-x_{2},1-x_{3},b_{2},b_{3})
,\\
F_{Ta3(4)}^{\left( d\right) } &=&32\pi
C_{F}M_{B}^{2}r_{k^*}^2\int_{0}^{1}dx_{2}dx_{3}\int_{0}^{\infty}
b_{2}db_{2}b_{3}db_{3}
\nonumber \\
&&\times
\left[(2-x_2)(\Phi^{v}_{k^*}(x_2)\Phi^{a}_{k^*}(x_3)-\Phi^{a}_{k^*}(x_2)\Phi^{v}_{k^*}(x_3))
\right.
\nonumber \\
&&\left.
+x_2(\Phi^{a}_{k^*}(x_2)\Phi^{a}_{k^*}(x_3)-\Phi^{v}_{k^*}(x_2)\Phi^{v}_{k^*}(x_3))\right]
\nonumber\\
&&\times
E_{a3(4)}^{(d)}(t_{ad}^{(1)})h_{a}(1-x_{2},1-x_{3},b_{2},b_{3})
,\label{fta3}
\end{eqnarray}
\begin{eqnarray}
 F_{Ta5}^{(d)} &=&-32\pi
C_{F}M_{B}^{2}r_{k^*}^2\int_{0}^{1}dx_{2}dx_{3}\int_{0}^{\infty}
b_{2}db_{2}b_{3}db_{3}
\nonumber \\
&&\times
\left[(2-x_2)(\Phi^{v}_{k^*}(x_2)\Phi^{a}_{k^*}(x_3)-\Phi^{a}_{k^*}(x_2)\Phi^{v}_{k^*}(x_3))
\right.
\nonumber \\
&&\left.
+x_2(\Phi^{a}_{k^*}(x_2)\Phi^{a}_{k^*}(x_3)-\Phi^{v}_{k^*}(x_2)\Phi^{v}_{k^*}(x_3))\right]
\nonumber\\
&&\times
E_{a5}^{(d)}(t_{ad}^{(1)})h_{a}(1-x_{2},1-x_{3},b_{2},b_{3})
,\label{fta5}\\
F_{La6}^{\left( d\right) } &=&32\pi
C_{F}M_{B}^{2}r_{k^*}\int_{0}^{1}dx_{2}dx_{3}\int_{0}^{\infty}
b_{2}db_{2}b_{3}db_{3}
\nonumber \\
&& \times \left[(1-x_2)(\Phi^s_{k^*}(x_2)+
\Phi^{t}_{k^*}(x_2))\Phi_{k^*}(x_3)+2\Phi_{k^*}(x_2)
\Phi^s_{k^*}(x_3) \right] \nonumber\\
&&
 \times E_{a6}^{(d)}(t^{(1)}_{ad})h_{a}(1-x_{2},1-x_{3},b_{2},b_{3})
,
\\
F_{Na6}^{\left( d\right) } &=&32\pi
C_{F}M_{B}^{2}r_{K^*}\int_{0}^{1}dx_{2}dx_{3}\int_{0}^{\infty}
b_{2}db_{2}b_{3}db_{3}
\nonumber \\
&& \times (\Phi^{T}_{K^*}(x_2)
\left(\Phi^v_{K^*}(x_{3})-\Phi^{a}_{K^*}(x_3)\right)
E_{a6}^{(d)}(t^{(1)}_{ad})h_{a}(1-x_2,1-x_3,b_2,b_3) ,\\
F_{Ta6}^{( d) }&=&2F_{Na6}^{( q) },\\
 F_{Ta3}^{\left(s\right) }
&=&32\pi
C_{F}M_{B}^{2}r_{K^*}^2\int_{0}^{1}dx_{2}dx_{3}\int_{0}^{\infty}
b_{2}db_{2}b_{3}db_{3}
\nonumber \\
&&\times
\left[(1-x_3)(\phi_{k^*}^a(x_2)\phi_{k^*}^a(x_3)-\phi_{k^*}^v(x_2)\phi_{k^*}^v(x_3))
\right.\nonumber \\
&&\left.
+(1+x_3)(\phi_{k^*}^v(x_2)\phi_{k^*}^a(x_3)-\phi_{k^*}^a(x_2)\phi_{k^*}^v(x_3))\right]
\nonumber\\
&&\times
E_{a3}^{(s)}(t_{as}^{(1)})h_{a}(x_{3},x_{2},b_{3},b_{2}),\\
F_{Ta5}^{(s)}&=&- F_{Ta3}^{\left(s\right)}.
\end{eqnarray}
The last expression of the factorizable amplitudes $F_{Ta5}^{(s)}$
doesn't really mean it equal to $F_{Ta3}^{(s)}$ but with the
evolution factor $E_{Ta3}^{(s)}$ replaced by $E_{Ta5}^{(s)}$ and
plus a factor $-1$ in the beginning. Other amplitudes which you
can not find in the upper formulas must be equal to zero.

The factors $E(t)$ contain the evolution from the $W$ boson mass
to the hard scales $t$ in the Wilson coefficients $a(t)$, and from
$t$ to the factorization scale $1/b$ in the Sudakov factors
$S(t)$:
\begin{eqnarray}
E_{e4}^{(q)}\left( t\right) &=&\alpha _{s}\left( t\right)
a_{4}^{(q)}(t) S_{B}\left( t\right)S_{K^{*}}\left( t\right)\;,
\nonumber \\
E_{ai}^{(q)}\left( t\right) &=&\alpha _{s}\left( t\right)
a_{i}^{(q)}(t) S_{K^{*}}(t)S_{K^{*}}(t) \;. \label{Eea}
\end{eqnarray}
The Wilson coefficients $a$ in the above formulas are given by
\begin{eqnarray}
a_{1}^{(q)} &=&C_{1}+\frac{C_{2}}{N_{c}}\;, \\
a_{2}^{(q)} &=&C_{2}+\frac{C_{1}}{N_{c}}\;, \\
a_{3}^{(q)} &=&\left( C_{3}+\frac{C_{4}}{N_{c}}\right)
+\frac{3}{2}e_{q}
\left(C_{9}+\frac{C_{10}}{N_{c}}\right) \;, \\
a_{4}^{(q)} &=&\left( C_{4}+\frac{C_{3}}{N_{c}}\right)
+\frac{3}{2}e_{q}
\left(C_{10}+\frac{C_{9}}{N_{c}}\right) \;, \\
a_{5}^{(q)} &=&\left( C_{5}+\frac{C_{6}}{N_{c}}\right)
+\frac{3}{2}e_{q}
\left(C_{7}+\frac{C_{8}}{N_{c}}\right) \;, \\
a_{6}^{(q)} &=&\left( C_{6}+\frac{C_{5}}{N_{c}}\right)
+\frac{3}{2}e_{q} \left(C_{8}+\frac{C_{7}}{N_{c}}\right).
\end{eqnarray}
$k_T$ resummation of large logarithmic corrections to the $B$,
$K^{*}$ and $K^{*}$ meson distribution amplitudes lead to the
exponentials $S_{B}$, $S_{K^{*}}$ and $S_{K^{*}}$, respectively.
\begin{eqnarray}
S_{B}(t)&=&\exp\left[-s(x_{1}P_{1}^{+},b_{1})
-2\int_{1/b_{1}}^{t}\frac{d{\bar{\mu}}} {\bar{\mu}} \gamma (\alpha
_{s}({\bar{\mu}}^2))\right]\;,
\nonumber \\
S_{K^{*} }(t)&=&\exp\left[-s(x_{2}P_{2}^{+},b_{2})
-s((1-x_{2})P_{2}^{+},b_{2})
-2\int_{1/b_{2}}^{t}\frac{d{\bar{\mu}}}{\bar{\mu}} \gamma (\alpha
_{s}({\bar{\mu}}^2))\right]\;,
\nonumber \\
S_{K^{*}}(t)&=&\exp\left[-s(x_{3}P_{3}^{-},b_{3})
-s((1-x_{3})P_{3}^{-},b_{3})
-2\int_{1/b_{3}}^{t}\frac{d{\bar{\mu}}}{\bar{\mu}} \gamma
(\alpha_{s}({\bar{\mu}}^2))\right]\;, \label{sbk}
\end{eqnarray}
 The
variables $b_{1}$, $b_{2}$, and $b_{3}$, conjugate to the parton
transverse momenta $k_{1T}$, $k_{2T}$, and $k_{3T}$, represent the
transverse extents of the $B$, $K^{*}$ and $K^{*}$ mesons,
respectively. The quark anomalous dimension
$\gamma=-{\alpha_s}/{\pi}$ and the so-called Sudakov factor
$s(Q,b)$ is expressed as
\begin{eqnarray}
  s(Q,b) &=& \int_{1/b}^Q \!\! \frac{d\mu'}{\mu'} \left[
 \left\{ \frac{2}{3}(2 \gamma_E - 1 - \log 2) + C_F \log \frac{Q}{\mu'}
 \right\} \frac{\alpha_s(\mu')}{\pi} \right. \nonumber \\
& &  \left.+ \left\{ \frac{67}{9} - \frac{\pi^2}{3} -
\frac{10}{27} n_f
 + \frac{2}{3} \beta_0 \log \frac{\gamma_E}{2} \right\}
 \left( \frac{\alpha_s(\mu')}{\pi} \right)^2 \log \frac{Q}{\mu'}
 \right].
 \label{eq:SudakovExpress}
\end{eqnarray}

 The above Sudakov exponentials decrease fast in the
large $b$ region, such that the $B\to K^{*}K^{*}$ hard amplitudes
remain sufficiently perturbative in the end-point region.

The hard functions $h$'s are
\begin{eqnarray}
h_{e}(x_{1},x_{2},b_{1},b_{2}) &=&K_{0}\left( \sqrt{x_{1}x_{2}}
M_{B}b_{1}\right)S_t(x_2)
\nonumber \\
&&\times \left[ \theta (b_{1}-b_{2})K_{0}\left( \sqrt{x_{2}}
M_{B}b_{1}\right) I_{0}\left( \sqrt{x_{2}}M_{B}b_{2}\right)
\right.
\nonumber \\
&&\left. +\theta (b_{2}-b_{1})K_{0}\left(
\sqrt{x_{2}}M_{B}b_{2}\right) I_{0}\left(
\sqrt{x_{2}}M_{B}b_{1}\right) \right] \;,
\label{he} \\
h_{a}(x_{2},x_{3},b_{2},b_{3}) &=&\left( \frac{i\pi
}{2}\right)^{2} H_{0}^{(1)}\left(
\sqrt{x_{2}x_{3}}M_{B}b_{3}\right)S_t(x_3)
\nonumber \\
&&\times \left[ \theta (b_{2}-b_{3})H_{0}^{(1)}\left( \sqrt{x_{2}}
M_{B}b_{2}\right) J_{0}\left( \sqrt{x_{2}}M_{B}b_{3}\right)
\right.
\nonumber \\
&&\left. +\theta (b_{3}-b_{2})H_{0}^{(1)}\left( \sqrt{x_{2}}
M_{B}b_{3}\right) J_{0}\left( \sqrt{x_{2}}M_{B}b_{2}\right)
\right] \;. \label{ha}
\end{eqnarray}
We have proposed the parametrization for the evolution function
$S_{t}(x)$ from threshold resummation\cite{threshold}.
\begin{eqnarray}
S_t(x)=\frac{2^{1+2c}\Gamma(3/2+c)}{\sqrt{\pi}\Gamma(1+c)}
[x(1-x)]^c\;. \label{str}
\end{eqnarray}
where the parameter $c$ is chosen as $c=0.4$ for the $B\to K^{*}
K^{*}$ decays. This factor modifies the end-point behavior of the
meson distribution amplitudes, rendering them vanish faster at
$x\to 0$. Threshold resummation for nonfactorizable diagrams is
weaker and negligible. $K_0, I_0, H_0$ and $J_0$ are the Bessel
functions.

The hard scales $t$ are chosen as the maxima of the virtualities
of the internal particles involved in the hard amplitudes,
including $1/b_{i}$:
\begin{eqnarray}
t_{e}^{(1)} &=&{\rm max}(\sqrt{x_{2}}M_{B},1/b_{1},1/b_{2})\;,
\nonumber \\
t_{e}^{(2)} &=&{\rm max}(\sqrt{x_{1}}M_{B},1/b_{1},1/b_{2})\;,
\nonumber \\
t_{ad(u)}^{(1)} &=&{\rm
max}(\sqrt{1-x_{2}}M_{B},1/b_{2},1/b_{3})\;,
\nonumber \\
t_{as}^{(1)} &=&{\rm
max}(\sqrt{x_{3}}M_{B},1/b_{2},1/b_{3})\;.\label{tfact}
\end{eqnarray}

\section{Nonfactorization formulas}
The nonfactorizable amplitudes depending on kinematic variables of
all the three mesons, are written as
\begin{eqnarray}
{\cal M}_{Le3} &=&16\pi C_{F}M_{B}^{2}\sqrt{2N_{c}}%
\int_{0}^{1}d[x]\int_{0}^{\infty }b_{1}db_{1}b_{2}db_{2} \Phi
_{B}(x_{1},b_{1})
\nonumber \\
&& \left\{\Phi _{K^* }(x_{3}) \left[ x_{3}\Phi _{K^*} ( x_{2} ) +
x_2 r_{K^*} (\Phi^{t}_{K^*}(x_2)-\Phi^{s}_{K^*}(x_2)) \right]
 E_{e3}^{\prime}(t_d^{(1)})h_d^{(1)}(x_1,x_2,x_3,b_1,b_2)\right.
\nonumber \\&& \left.+\Phi _{K^* }(x_{3}) \left[ (-1-x_{2}+x_3)
\Phi_{K^*}( x_{2} ) +x_2
r_{K^*}(\Phi^{t}_{K^*}(x_2)+\Phi^{s}_{K^*}(x_2)) \right]\right.
\nonumber \\&& \left.
 \times E_{e3}^{\prime}(t^{(2)}_d)
h_d^{(2)}(x_1,x_2,x_3,b_1,b_2) \right\},
\\
{\cal M}_{Ne3} &=&16\pi C_{F}M_{B}^{2}\sqrt{2N_{c}}r_{K^*}
\int_{0}^{1}d[x]\int_{0}^{\infty }b_{1}db_{1}b_{2}db_{2} \Phi
_{B}(x_{1},b_{1})
\nonumber \\
&& \left\{ x_3\Phi^{T}_{K^*}(x_2)
(\Phi^v_{K^*}(x_3)-\Phi^{a}_{K^*}(x_3))
E_{e3}^{\prime}(t^{(1)}_d)h_d^{(1)}(x_1,x_2,x_3,b_1,b_2)\right.
\nonumber \\
&&   +
\left[(1-x_3)\Phi^{T}_{\phi}(x_2)(\Phi^{v}_{K^*}(x_3)-\Phi^{a}_{\phi}(x_3))
+2r_{k^*}(1+x_2-x_3)\right.
\nonumber \\
&&  \left.\left. \times(\Phi^{a}_{K^*}(x_2)\Phi^{a}_{K^*}(x_3)
-\Phi^{v}_{K^*}(x_2)\Phi^{v}_{K^*}(x_3))\right]
 E_{e3}^{\prime}(t^{(2)}_d)
h_d^{(2)}(x_1,x_2,x_3,b_1,b_2) \right\}\;,
\\
{\cal M}_{Te3} &=&32\pi C_{F}M_{B}^{2}\sqrt{2N_{c}}r_{K^*}
\int_{0}^{1}d[x]\int_{0}^{\infty }b_{1}db_{1}b_{2}db_{2} \Phi
_{B}(x_{1},b_{1})
\nonumber \\
&& \left\{ x_3\Phi^{T}_{K^*}(x_2)
(\Phi^v_{K^*}(x_3)-\Phi^{a}_{K^*}(x_3))
E_{e3}^{\prime}(t^{(1)}_d)h_d^{(1)}(x_1,x_2,x_3,b_1,b_2)\right.
\nonumber \\
&&   +
\left[(1-x_3)\Phi^{T}_{\phi}(x_2)(\Phi^{v}_{K^*}(x_3)-\Phi^{a}_{\phi}(x_3))
+2r_{k^*}(1+x_2-x_3)\right.
\nonumber \\
&&  \left.\left. \times(\Phi^{v}_{K^*}(x_2)\Phi^{a}_{K^*}(x_3)
-\Phi^{a}_{K^*}(x_2)\Phi^{v}_{K^*}(x_3))\right]
 E_{e3}^{\prime}(t^{(2)}_d)
h_d^{(2)}(x_1,x_2,x_3,b_1,b_2) \right\}\;, \\
{\cal M}_{Le5} &=&16\pi C_{F}M_{B}^{2}\sqrt{2N_{c}}r_{K^*}
\int_{0}^{1}d[x]\int_{0}^{\infty }b_{1}db_{1}b_{2}db_{2}
\Phi_{B}(x_{1},b_{1})
\nonumber \\
&& \left \{ \left[ x_3  \Phi_{K^*}(x_2)(
\Phi^{s}_{K^*}(x_3)-\Phi^{t}_{\phi}(x_3))
+r_{K^*}(x_2+x_3)(\Phi^{s}_{K^*}(x_2)
\Phi^{s}_{K^*}(x_3)+\Phi^{t}_{K^*}(x_2)
\Phi^{t}_{K^*}(x_3))\right.\right.
\nonumber\\
&& \left. +r_{K^*}(x_2-x_3)(\Phi^{s}_{K^*}(x_2)
\Phi^{t}_{K^*}(x_3)+\Phi^{t}_{K^*}(x_2)
\Phi^{s}_{K^*}(x_3))\right]
 E^{\prime}_{e5}(t^{(1)}_{d})
h_{d}^{(1)}(x_{1},x_{2},x_{3},b_{1},b_{2})
\nonumber \\
&& +\left[(-1+x_3)  \Phi_{K^*}(x_2)(
\Phi^{s}_{K^*}(x_3)+\Phi^{t}_{\phi}(x_3))\right.
\nonumber\\
&&   +r_{K^*}(-1-x_2+x_3)(\Phi^{s}_{K^*}(x_2)
\Phi^{s}_{K^*}(x_3)-\Phi^{t}_{K^*}(x_2)
\Phi^{t}_{K^*}(x_3))
\nonumber\\
&& \left.+r_{K^*}(-1+x_2+x_3)(\Phi^{s}_{K^*}(x_2)
\Phi^{t}_{K^*}(x_3)-\Phi^{t}_{K^*}(x_2)
\Phi^{s}_{K^*}(x_3))\right]
\nonumber\\
&& \left.\times
 E_{e5}^{\prime}(t^{(2)}_d)h_{d}^{(2)}(
x_{1},x_{2},x_{3},b_{1},b_{2}) \right\}\;,
\end{eqnarray}
\begin{eqnarray}
{\cal M}_{Ne5} &=&16\pi C_{F}M_{B}^{2}\sqrt{2N_{c}}r_{k^*}
\int_{0}^{1}d[x]\int_{0}^{\infty }b_{1}db_{1}b_{2}db_{2} \Phi
_{B}(x_{1},b_{1}) \nonumber \\ &&
 \left\{ \left[ x_{2}
 (\Phi^{a}_{k^*}(x_2)-\Phi^{v}_{k^*}(x_2))\Phi^{T}_{\phi}(x_3)
+r_{k^*} (x_2+x_3)\Phi^T_{k^*}(x_2)\Phi^T_{k^*}(x_3)
\right]\right.
\nonumber\\
&&\times E^{\prime}_{e5}(t^{(1)}_{d})
h_{d}^{(1)}(x_{1},x_{2},x_{3},b_{1},b_{2})\nonumber\\
&&+  \left[ x_{2}
(\Phi^{a}_{k^*}(x_2)-\Phi^{v}_{k^*}(x_2))\Phi^{T}_{k^*}(x_3)
+r_{K^*} (1+x_2-x_3)\Phi^T_{k^*}(x_2)\Phi^T_{k^*}(x_3) \right]\nonumber\\
&&\left.
 \times E^{\prime}_{e5}(t^{(2)}_{d})h_{d}^{(2)}(x_{1},x_{2},x_{3},b_{1},b_{2})
\right\}\;,\\
 {\cal M}_{Te5} &=&32\pi
C_{F}M_{B}^{2}\sqrt{2N_{c}}r_{k^*}
\int_{0}^{1}d[x]\int_{0}^{\infty }b_{1}db_{1}b_{2}db_{2} \Phi
_{B}(x_{1},b_{1}) \nonumber \\ &&
 \left\{ \left[ x_{2}
 (\Phi^{a}_{k^*}(x_2)-\Phi^{v}_{k^*}(x_2))\Phi^{T}_{\phi}(x_3)
+r_{k^*} (x_2-x_3)\Phi^T_{k^*}(x_2)\Phi^T_{k^*}(x_3)
\right]\right. \nonumber\\&&
 \times E^{\prime}_{e5}(t^{(1)}_{d})
h_{d}^{(1)}(x_{1},x_{2},x_{3},b_{1},b_{2})\nonumber\\
&&+  \left[ x_{2}
(\Phi^{a}_{k^*}(x_2)-\Phi^{v}_{k^*}(x_2))\Phi^{T}_{k^*}(x_3)
+r_{K^*} (-1+x_2+x_3)\Phi^T_{k^*}(x_2)\Phi^T_{k^*}(x_3) \right]\nonumber\\
&&\left.
 \times E^{\prime}_{e5}(t^{(2)}_{d})h_{d}^{(2)}(x_{1},x_{2},x_{3},b_{1},b_{2})
\right\}\;,\\
 {\cal M}_{La3(4)}^{( d) } &=&16\pi
C_{F}M_{B}^{2}\sqrt{2N_{c}} \int_{0}^{1}d[x]\int_{0}^{\infty
}b_{1}db_{1}b_{2}db_{2} \Phi _{B}(x_{1},b_{1})
\nonumber \\
&& \left\{ \left[ (-1+x_3) \Phi_{k^*}(x_2)\Phi_{k^*}(x_3) \right.
\right.
\nonumber\\
&&+r_{k^*}^2 \left(-4\Phi^{s}_{k^*}(x_2)\Phi^{s}_{k^*}(x_3)
+(x_2+x_3)(\Phi^{s}_{k^*}(x_2)\Phi^{s}_{k^*}(x_3)+\Phi^{t}_{k^*}(x_2)\Phi^{t}_{k^*}(x_3))\right.
\nonumber\\
&&\left.\left.+(x_2-x_3)(\Phi^{s}_{k^*}(x_2)\Phi^{t}_{k^*}(x_3)+\Phi^{t}_{k^*}(x_2)\Phi^{s}_{k^*}(x_3))\right)
\right]
\nonumber\\
&&  \times E^{(d)\prime}_{a3(4)}(t^{(1)}_{f})
h_{fd}^{(1)}(x_{1},x_{2},x_{3},b_{1},b_{2})
\nonumber \\
&& +\left[ (1-x_2) \Phi_{k^*}(x_2)\Phi_{k^*}(x_3) \right.
\nonumber\\
&&+r_{k^*}^2 \left(
(2-x_2-x_3)(\Phi^{s}_{k^*}(x_2)\Phi^{s}_{k^*}(x_3)+\Phi^{t}_{k^*}(x_2)\Phi^{t}_{k^*}(x_3))\right.
\nonumber\\
&&\left.\left.+(x_2-x_3)(\Phi^{s}_{k^*}(x_2)\Phi^{t}_{k^*}(x_3)+\Phi^{t}_{k^*}(x_2)\Phi^{s}_{k^*}(x_3))\right)
\right]
\nonumber\\
&& \left.\times E^{(d)\prime}_{a3(4)}(t^{(2)}_{d})
h_{fd}^{(2)}(x_{1},x_{2},x_{3},b_{1},b_{2}) \right\}\;,
\\
{\cal M}_{Na3(4)}^{(d) } &=&16\pi C_{F}M_{B}^{2}\sqrt{2N_{c}}
r_{K^*}^2 \int_{0}^{1}d[x]\int_{0}^{\infty }b_{1}db_{1}b_{2}db_{2}
\Phi _{B}(x_{1},b_{1}) \nonumber
\\&& \left\{ \left[2(\Phi^{a}_{K^*}(x_2)\Phi^{a}_{K^*}(x_3)-\Phi^{v}_{K^*}(x_2)\Phi^{v}_{K^*}(x_3))
+(x_2+x_3)\Phi^{T}_{K^*}(x_2)\Phi^{T}_{K^*}(x_3)\right]\right.
\nonumber \\
&&\times E^{(d)\prime}_{a3(4)}(t^{(1)}_{f})
h_{fd}^{(1)}(x_{1},x_{2},x_{3},b_{1},b_{2})
\nonumber \\
&& +\left.(2- x_2-x_3)\Phi^{T}_{K^*}(x_2)\Phi^{T}_{K^*}(x_3)
E^{(d)\prime}_{a3(4)}(t^{(2)}_{f})
h_{fd}^{(2)}(x_{1},x_{2},x_{3},b_{1},b_{2})\right\}\;,
\end{eqnarray}
\begin{eqnarray}
{\cal M}_{Ta3(4)}^{(d) } &=&32\pi C_{F}M_{B}^{2}\sqrt{2N_{c}}
r_{K^*}^2 \int_{0}^{1}d[x]\int_{0}^{\infty }b_{1}db_{1}b_{2}db_{2}
\Phi _{B}(x_{1},b_{1}) \nonumber
\\&& \left\{ \left[2(\Phi^{v}_{K^*}(x_2)\Phi^{a}_{K^*}(x_3)-\Phi^{a}_{K^*}(x_2)\Phi^{v}_{K^*}(x_3))
+(x_2-x_3)\Phi^{T}_{K^*}(x_2)\Phi^{T}_{K^*}(x_3)\right]\right.
\nonumber \\
&&\times E^{(d)\prime}_{a3}(t^{(1)}_{f})
h_{fd}^{(1)}(x_{1},x_{2},x_{3},b_{1},b_{2})
\nonumber \\
&& +\left.(x_3- x_2)\Phi^{T}_{K^*}(x_2)\Phi^{T}_{K^*}(x_3)
E^{(f)\prime}_{a3(4)}(t^{(2)}_{f})
h_{fd}^{(2)}(x_{1},x_{2},x_{3},b_{1},b_{2})\right\}\;,\\
 {\cal
M}_{La5}^{(d) } &=&16\pi C_{F}M_{B}^{2}\sqrt{2N_{c}}r_{K^*}
\int_{0}^{1}d[x]\int_{0}^{\infty }b_{1}db_{1}b_{2}db_{2} \Phi
_{B}(x_{1},b_{1})
\nonumber \\
&&\left\{ \left[ (1+x_2) (\Phi^{t}_{K^*}(x_2)-
\Phi^{s}_{K^*}(x_2))\Phi_{K^*}(x_3)+(1+x_{3})\Phi_{K^*}(x_2)(\Phi^{s}_{K^*}(x_3)
-\Phi^{t}_{K^*}(x_3))\right]\right.\nonumber \\
&&
 \times E^{(d)\prime}_{a5}(t^{(1)}_{f})
h_{fd}^{(1)}(x_{1},x_{2},x_{3},b_{1},b_{2})
\nonumber \\
&&  + \left[ (1-x_2) (\Phi^{t}_{K^*}(x_2)-
\Phi^{s}_{K^*}(x_2))\Phi_{K^*}(x_3)+(1-x_{3})\Phi_{K^*}(x_2)(\Phi^{s}_{K^*}(x_3)
-\Phi^{t}_{K^*}(x_3))\right]\nonumber \\
&&\left.\times
E^{(d)\prime}_{a5}(t^{(2)}_{f})h_{fd}^{(2)}(x_{1},x_{2},x_{3},b_{1},b_{2})
\right\}\;,\\
 {\cal M}_{Na5}^{(d) } &=&16\pi
C_{F}M_{B}^{2}\sqrt{2N_{c}}r_{K^*}
\int_{0}^{1}d[x]\int_{0}^{\infty }b_{1}db_{1}b_{2}db_{2} \Phi
_{B}(x_{1},b_{1})
\nonumber \\
&&\left\{\left[(1+x_2)(\Phi^{a}_{K^*}(x_2)
-\Phi^{v}_{K^*}(x_2))\Phi^T_{K^*}(x_3)+
 (1+x_3) \Phi^T_{K^*}(x_2)
(\Phi^{v}_{K^*}(x_3)-\Phi^{a}_{K^*}(x_3))\right]\right.
\nonumber \\
&&\left.\times
 E^{(d)\prime}_{a5}(t^{(1)}_{f})
h_{fd}^{(1)}(x_{1},x_{2},x_{3},b_{1},b_{2})\right.
\nonumber \\
&& +\left[(1-x_2)(\Phi^{a}_{K^*}(x_2)
-\Phi^{v}_{K^*}(x_2))\Phi^T_{K^*}(x_3)+
 (1-x_3) \Phi^T_{K^*}(x_2)
(\Phi^{v}_{K^*}(x_3)-\Phi^{a}_{K^*}(x_3))\right]
\nonumber \\
&& \times \left. E^{(d)\prime}_{a5}(t^{(2)}_{f})
h_{fd}^{(2)}(x_{1},x_{2},x_{3},b_{1},b_{2})\right\}\;,
\\
{\cal M}_{Ta5}^{(d) }&=& 2 {\cal M}_{Na5}^{(d) }\;.\\
 {\cal M}_{L6}^{( d) } &=&16\pi C_{F}M_{B}^{2}\sqrt{2N_{c}}
\int_{0}^{1}d[x]\int_{0}^{\infty }b_{1}db_{1}b_{2}db_{2} \Phi
_{B}(x_{1},b_{1})
\nonumber \\
&& \left\{ \left[ (-1+x_2) \Phi_{k^*}(x_2)\Phi_{k^*}(x_3) \right.
\right.
\nonumber\\
&&+r_{k^*}^2 \left(-4\Phi^{s}_{k^*}(x_2)\Phi^{s}_{k^*}(x_3)
+(x_2+x_3)(\Phi^{s}_{k^*}(x_2)\Phi^{s}_{k^*}(x_3)+\Phi^{t}_{k^*}(x_2)\Phi^{t}_{k^*}(x_3))\right.
\nonumber\\
&&\left.\left.+(x_3-x_2)(\Phi^{s}_{k^*}(x_2)\Phi^{t}_{k^*}(x_3)+\Phi^{t}_{k^*}(x_2)\Phi^{s}_{k^*}(x_3))\right)
\right]
\nonumber\\
&&  \times E^{(d)\prime}_{a6}(t^{(1)}_{f})
h_{fd}^{(1)}(x_{1},x_{2},x_{3},b_{1},b_{2})
\nonumber \\
&& +\left[ (1-x_3) \Phi_{k^*}(x_2)\Phi_{k^*}(x_3) \right.
\nonumber\\
&&+r_{k^*}^2 \left(
(2-x_2-x_3)(\Phi^{s}_{k^*}(x_2)\Phi^{s}_{k^*}(x_3)+\Phi^{t}_{k^*}(x_2)\Phi^{t}_{k^*}(x_3))\right.
\nonumber\\
&&\left.\left.+(x_3-x_2)(\Phi^{s}_{k^*}(x_2)\Phi^{t}_{k^*}(x_3)+\Phi^{t}_{k^*}(x_2)\Phi^{s}_{k^*}(x_3))\right)
\right]
\nonumber\\
&& \left.\times E^{(d)\prime}_{a6}(t^{(2)}_{d})
h_{fd}^{(2)}(x_{1},x_{2},x_{3},b_{1},b_{2}) \right\}\;,
\\
{\cal M}_{Na6}^{(d) } &=&{\cal M}_{Na3}^{(d) }
\ \\
{\cal M}_{Ta6}^{(d) } &=&-{\cal M}_{Ta3}^{(d) }
\end{eqnarray}
\begin{eqnarray}
{\cal M}_{La4}^{( s) } &=&-16\pi C_{F}M_{B}^{2}\sqrt{2N_{c}}
\int_{0}^{1}d[x]\int_{0}^{\infty }b_{1}db_{1}b_{2}db_{2} \Phi
_{B}(x_{1},b_{1})
\nonumber \\
&& \left\{ \left[ x_2 \Phi_{k^*}(x_2)\Phi_{k^*}(x_3)+r_{k^*}^2
\left(2(\Phi^{s}_{k^*}(x_2)\Phi^{s}_{k^*}(x_3)-\Phi^{t}_{k^*}(x_2)\Phi^{t}_{k^*}(x_3))
\right. \right.\right.
\nonumber\\
&&\left.
+(x_2+x_3)(\Phi^{s}_{k^*}(x_2)\Phi^{s}_{k^*}(x_3)+\Phi^{t}_{k^*}(x_2)\Phi^{t}_{k^*}(x_3))\right.
\nonumber\\
&&\left.\left.+(x_2-x_3)(\Phi^{s}_{k^*}(x_2)\Phi^{t}_{k^*}(x_3)+\Phi^{t}_{k^*}(x_2)\Phi^{s}_{k^*}(x_3))\right)
\right]
\nonumber\\
&&  \times E^{(s)\prime}_{a4}(t^{(1)}_{f})
h_{fs}^{(1)}(x_{1},x_{2},x_{3},b_{1},b_{2})
\nonumber \\
&& +\left[ -x_3 \Phi_{k^*}(x_2)\Phi_{k^*}(x_3) +r_{k^*}^2 \left(
-(x_2+x_3)(\Phi^{s}_{k^*}(x_2)\Phi^{s}_{k^*}(x_3)+\Phi^{t}_{k^*}(x_2)\Phi^{t}_{k^*}(x_3))\right.\right.
\nonumber\\
&&\left.\left.+(x_2-x_3)(\Phi^{s}_{k^*}(x_2)\Phi^{t}_{k^*}(x_3)+\Phi^{t}_{k^*}(x_2)\Phi^{s}_{k^*}(x_3))\right)
\right]
\nonumber\\
&& \left.\times E^{(s)\prime}_{a4}(t^{(2)}_{d})
h_{fs}^{(2)}(x_{1},x_{2},x_{3},b_{1},b_{2}) \right\}\;,
\\
{\cal M}_{Na4}^{(s) } &=&16\pi C_{F}M_{B}^{2}\sqrt{2N_{c}}
r_{K^*}^2 \int_{0}^{1}d[x]\int_{0}^{\infty }b_{1}db_{1}b_{2}db_{2}
\Phi _{B}(x_{1},b_{1}) \nonumber
\\&& \left\{ \left[2(\Phi^{a}_{K^*}(x_2)\Phi^{a}_{K^*}(x_3)-\Phi^{v}_{K^*}(x_2)\Phi^{v}_{K^*}(x_3))
+(2-x_2-x_3)\Phi^{T}_{K^*}(x_2)\Phi^{T}_{K^*}(x_3)\right]\right.
\nonumber \\
&&\times E^{(s)\prime}_{a4}(t^{(1)}_{f})
h_{fs}^{(1)}(x_{1},x_{2},x_{3},b_{1},b_{2})
\nonumber \\
&& +\left.(x_2+x_3)\Phi^{T}_{K^*}(x_2)\Phi^{T}_{K^*}(x_3)
E^{(s)\prime}_{a3(4)}(t^{(2)}_{f})
h_{fs}^{(2)}(x_{1},x_{2},x_{3},b_{1},b_{2})\right\}\;,
\\
{\cal M}_{Ta4}^{(s) } &=&32\pi C_{F}M_{B}^{2}\sqrt{2N_{c}}
r_{K^*}^2 \int_{0}^{1}d[x]\int_{0}^{\infty }b_{1}db_{1}b_{2}db_{2}
\Phi _{B}(x_{1},b_{1}) \nonumber
\\&& \left\{ \left[2(\Phi^{v}_{K^*}(x_2)\Phi^{a}_{K^*}(x_3)-\Phi^{a}_{K^*}(x_2)\Phi^{v}_{K^*}(x_3))
+(x_2-x_3)\Phi^{T}_{K^*}(x_2)\Phi^{T}_{K^*}(x_3)\right]\right.
\nonumber \\
&&\times E^{(s)\prime}_{a4}(t^{(1)}_{f})
h_{fs}^{(1)}(x_{1},x_{2},x_{3},b_{1},b_{2})
\nonumber \\
&& +\left.(x_3-x_2)\Phi^{T}_{K^*}(x_2)\Phi^{T}_{K^*}(x_3)
E^{(s)\prime}_{a3(4)}(t^{(2)}_{f})
h_{fs}^{(2)}(x_{1},x_{2},x_{3},b_{1},b_{2})\right\}\;,
\\
{\cal M}_{L6}^{( s) } &=&-16\pi C_{F}M_{B}^{2}\sqrt{2N_{c}}
\int_{0}^{1}d[x]\int_{0}^{\infty }b_{1}db_{1}b_{2}db_{2} \Phi
_{B}(x_{1},b_{1})
\nonumber \\
&& \left\{ \left[ x_3 \Phi_{k^*}(x_2)\Phi_{k^*}(x_3) +r_{k^*}^2
\left(2(\Phi^{s}_{k^*}(x_2)\Phi^{s}_{k^*}(x_3)-\Phi^{t}_{k^*}(x_2)\Phi^{t}_{k^*}(x_3))
\right. \right.\right.
\nonumber\\
&&
+(x_2+x_3)(\Phi^{s}_{k^*}(x_2)\Phi^{s}_{k^*}(x_3)+\Phi^{t}_{k^*}(x_2)\Phi^{t}_{k^*}(x_3))
\nonumber\\
&&\left.\left.+(x_3-x_2)(\Phi^{s}_{k^*}(x_2)\Phi^{t}_{k^*}(x_3)+\Phi^{t}_{k^*}(x_2)\Phi^{s}_{k^*}(x_3))\right)
\right]
\nonumber\\
&&  \times E^{(s)\prime}_{a6}(t^{(1)}_{f})
h_{fs}^{(1)}(x_{1},x_{2},x_{3},b_{1},b_{2})
\nonumber \\
&& +\left[ -x_2 \Phi_{k^*}(x_2)\Phi_{k^*}(x_3)+r_{k^*}^2 \left(
-(x_2+x_3)(\Phi^{s}_{k^*}(x_2)\Phi^{s}_{k^*}(x_3)
+\Phi^{t}_{k^*}(x_2)\Phi^{t}_{k^*}(x_3)) \nonumber\right.\right.
\\
&&\left.\left.+(x_3-x_2)(\Phi^{s}_{k^*}(x_2)\Phi^{t}_{k^*}(x_3)+\Phi^{t}_{k^*}(x_2)\Phi^{s}_{k^*}(x_3))\right)
\right]
\nonumber\\
&& \left.\times E^{(s)\prime}_{a6}(t^{(2)}_{d})
h_{fs}^{(2)}(x_{1},x_{2},x_{3},b_{1},b_{2}) \right\}\;,
\\{\cal M}_{N6}^{( s) } &=&{\cal M}_{N4}^{( s) },
\\
{\cal M}_{T6}^{( s) } &=&-{\cal M}_{T4}^{( s) }.
\end{eqnarray}
The expressions of the nonfactorizable amplitudes ${\cal M}_{Ha}$
and ${\cal M}_{He4}$ are the same as ${\cal M}_{Ha3}^{(q)}$ and
${\cal M}_{He3}^{(q)}$ but with the evolution factors
$E_{a3}^{(q)\prime}$ and $E_{e3}^{(q)\prime}$ replaced by
$E_{a1}^{(q)\prime}$ and $E_{e4}^{(q)\prime}$, respectively.

The evolution factors are given by
\begin{eqnarray}
E_{ei}^{\left( q\right) \prime }\left( t\right) &=&\alpha
_{s}\left( t\right) a_{i}^{\left( q\right)\prime}(t)S\left(
t^{\prime}\right)|_{b_{3}=b_{1}}\;,
 \nonumber \\
E_{ai}^{\left( q\right) \prime }\left( t\right) &=&\alpha
_{s}\left( t\right) a_{i}^{\left( q\right)\prime}(t)S\left(
t^{\prime}\right)|_{b_{3}=b_{2}}\;, \label{Eeaprim}
\end{eqnarray}
with the Sudakov factor $S=S_{B}S_{K^{*} }S_{K^{*}}$. The Wilson
coefficients $a$ appearing in the above formulas are
\begin{eqnarray*}
a_{1}^{\prime } &=&\frac{C_{1}}{N_{c}}\;, \\
a_{2}^{\prime }&=&\frac{C_{2}}{N_{c}}\;, \\
a_{3}^{(q)\prime} &=&\frac{1}{N_{c}}
\left(C_{3}+\frac{3}{2}e_{q}C_{9}\right)\;,\\
a_{4}^{(q)\prime}&=&\frac{1}{N_{c}}
\left( C_{4}+\frac{3}{2}e_{q}C_{10}\right)\;,\\
a_{5}^{(q)\prime}&=&\frac{1}{N_{c}}
\left( C_{5}+\frac{3}{2}e_{q}C_{7}\right) \;,\\
a_{6}^{(q)\prime} &=&\frac{1}{N_{c}}
\left(C_{6}+\frac{3}{2}e_{q}C_{8}\right) \;.
\end{eqnarray*}

The hard functions $h^{(j)}$, $j=1$ and 2, with $d$ stand for
emission and $f$ stand for annihilation, are written as
\begin{eqnarray}
h_{d}^{(j)}(x_1,x_2,x_3,b_1,b_3)  &
=&\left[\theta(b_1-b_3)K_0(DM_Bb_1) I_0(DM_Bb_3)\nonumber\right.\\
 &&\left. +\theta(b_3-b_1)K_0(DM_Bb_3) I_0(DM_Bb_1)\right]\nonumber \\
&&\times \left\{\begin{array}{ll} K_0(D_jM_Bb_3), \hspace{0.3cm}
\;\;\;\;\;\;\;\;\;\;\;\;\;\;D_j^2\geq 0,
\\ \frac{i\pi}{2}H_0^{(1)}(\sqrt{|D_j^2|}M_Bb_3),\ \ \ \ \ D_j^2\le0.
\end{array}\right.\label{hd}\\
h_{fq}^{(j)}(x_1,x_2,x_3,b_1,b_3) &=&\frac{i\pi }{2}\left[ \theta
(b_{1}-b_{3})H_{0}^{(1)} \left(F_{q}M_{B}b_{1}\right) J_{0}\left(
F_{q}M_{B}b_{3}\right) \right.
\nonumber \\
&&\quad \left. +\theta (b_{3}-b_{1})H_{0}^{(1)}\left(
F_qM_{B}b_{3}\right) J_{0}\left( F_qM_{B}b_{3}\right)\right]
\nonumber \\
&&\times \left\{\begin{array}{ll} K_{0}(F_{jq}M_{B}b_1),
\hspace{0.3cm};\;\;\;\;\;\;\;\;\;\;\;\; F^2_{jq} \geq 0,
\\ \frac{i\pi}{2}H_{0}^{(1)}
\left(\sqrt{|F_{jq}^{2}|}M_{B}b_1\right),\hspace{0.3cm}  F^2_{jq}
\leq 0.\end{array}\right.\label{hf}
\end{eqnarray}
with the variables,
\begin{eqnarray}
D^{2} &=& x_{1}x_{2}\;,
\nonumber \\
D_{1}^{2} &=&x_{2}(x_{1}-x_{3})\;,
\nonumber \\
D_{2}^{2} &=&-x_{2}(1-x_1-x_{3})\;,\nonumber  \\
F^{2}_{d(u)} &=&(1- x_{2}) \left( 1-x_{3}\right) \;,
\nonumber \\
F_{1d(u)}^{2} &=&1-x_{2}\left(x_{3}-x_1\right)\;,
\nonumber \\
F_{2d(u)}^{2} &=&(1-x_{2})\left(x_{1}+x_{3}-1\right)\;,\nonumber \\
F^{2}_s &=& x_{2} x_{3} \;,
\nonumber \\
F_{1s}^{2} &=&1-(1-x_{2})\left(1-x_{1}-x_3\right)\;,
\nonumber \\
F_{2s}^{2} &=&x_{2}\left(x_{1}-x_{3}\right)\;. \label{DF}
\end{eqnarray}
The hard scales $t^{(j)}$ are chosen as
\begin{eqnarray}
t_{d}^{(1){\prime}} &=&{\rm
max}\left(DM_{B},\sqrt{|D_{1}^{2}|}M_{B},1/b_{1},
1/b_{3}\right)\;,
\nonumber \\
t_{d}^{(2){\prime}} &=&{\rm
max}\left(DM_{B},\sqrt{|D_{2}^{2}|}M_{B},1/b_{1},
1/b_{3}\right)\;,
\nonumber \\
t_{fq}^{(1){\prime}} &=&{\rm
max}\left(F_qM_{B},\sqrt{|F_{1q}^{2}|}M_{B},1/b_{1},
1/b_{3}\right)\;,
\nonumber \\
t_{fq}^{(2){\prime}} &=&{\rm
max}\left(F_qM_{B},\sqrt{|F_{2q}^{2}|}M_{B},1/b_{1},
1/b_{3}\right)\;.\label{tnonfact}
\end{eqnarray}

\end{appendix}

\end{document}